\documentclass[a4paper,11pt]{article}
\pdfoutput=1
\usepackage{jheppub}
\usepackage[T1]{fontenc}
\usepackage[latin1]{inputenc}
\usepackage[english]{babel}
\usepackage{amsthm}
\usepackage{empheq}
\usepackage{mathdots}
\DeclareMathOperator{\Tr}{Tr}

\DeclareMathOperator{\adj}{ad}

\usepackage{dsfont}
\usepackage{pifont}
\newcommand{\cmark}{\ding{51}}
\newcommand{\xmark}{\ding{55}}
\usepackage[textsize=small,shadow]{todonotes}
\usepackage{ulem}
\hyphenchar\font=\string"7F

\title{\boldmath Generalized Metric Formulation of \\ Double Field Theory on Group Manifolds}
\preprint{LMU-ASC 03/15\\MPP-2015-14\\CERN-PH-TH-2015-020}

\author[a]{Ralph Blumenhagen,}
\emailAdd{blumenha@mpp.mpg.de}
\author[b]{Pascal du Bosque,}
\emailAdd{p.bosque@physik.lmu.de}
\author[a]{Falk Hassler,}
\emailAdd{fhassler@mpp.mpg.de}
\author[a,b,c]{and Dieter L\" ust}
\emailAdd{dieter.luest@lmu.de}

\affiliation[a]{Max-Planck-Institut f\"ur Physik\\
F\"ohringer Ring 6, 80805 M\"unchen, Germany}
\affiliation[b]{Arnold-Sommerfeld-Center f\"ur Theoretische Physik\\
Department f\"ur Physik, Ludwig-Maximilians-Universit\"at M\"unchen\\
Theresienstra\ss e 37, 80333 M\"unchen, Germany}
\affiliation[c]{CERN, PH-TH\\
1211 Geneva 23, Switzerland}

\abstract{We rewrite the recently derived cubic action of Double Field Theory on group manifolds \cite{Blumenhagen:2014gva} in terms of a generalized metric and extrapolate it to all orders in the fields. For the resulting action, we derive the field equations and state them in terms of a generalized curvature scalar and a generalized Ricci tensor. Compared to the generalized metric formulation of DFT derived from tori, all these quantities receive additional contributions related to the non-trivial background. It is shown that the action is invariant under its generalized diffeomorphisms and 2D-diffeomorphisms. Imposing additional constraints relating the background and fluctuations around it, the precise relation between the proposed generalized metric formulation of DFT${}_\mathrm{WZW}$ and of original DFT from tori is clarified. Furthermore, we show how to relate DFT${}_\mathrm{WZW}$ of the WZW background with  the flux formulation of original DFT.}

\begin{document}
\maketitle

\section{Introduction}
For several years, dualities have became a well established instrument to study fundamental aspects of string theory and the corresponding low energy effective field theories. Hence, it is not a surprise that there is a growing interest in a theory called Double Field Theory (DFT) \cite{Siegel:1993th,Hull:2009mi,Hull:2009zb,Hohm:2010jy,Hohm:2011ex,Aldazabal:2013sca,Berman:2013eva,Hohm:2013bwa} which makes abelian T-duality a manifest symmetry in the low energy description of closed string theory. To this end, it seizes the idea \cite{Tseytlin:1990nb,Siegel:1993th,Dabholkar:2002sy,Dabholkar:2005ve,Hull:2006va} to double the coordinates of the target space. Adding $D$ additional dual coordinates allows to take winding excitations of the closed string on a compact background into account. Exchanging winding and momentum excitations is the mechanism underpinning T-duality on a torus and thus the doubled target space of DFT permits to capture this mechanism through a global $O(D,D)$ symmetry. The doubling of the coordinates can be also viewed as introducing $D$ left-moving and $D$ right-moving closed string coordinates, where the ordinary and dual coordinates are just the sums and the differences of left- and right-moving
coordinates.

However, there are still conceptual questions about the current status of DFT. They are mainly triggered by the strong constraint which is required for a consistent low energy formulation. The strong constraint is a consequence of the toroidal background used in the original derivation \cite{Hull:2009mi} and it states that winding and momentum excitations in the same direction are not allowed. Violating the strong constraint, it is impossible to choose a torus radius in the corresponding direction to make all fields much lighter than the string scale. Either momentum or winding modes are heavier  than the first massive string excitations and spoil a consistent truncation. On the other hand, applying the strong constraint identifies DFT with the well studied NS/NS sector of SUGRA. Thus, except for an effective rewriting, it does not give any new physical insights. Moreover, such a rewriting is also available in terms of Hitchin's generalized complex geometry \cite{Hitchin:2004ut,Gualtieri:2003dx} which is an appropriate replacement for DFT in this case. The situation is more intriguing, but unfortunately also more speculative, if one weakens the strong constraint. In this case so called non-geometric backgrounds \cite{Hull:2004in,Dabholkar:2005ve,Hull:2006va,Hull:2009sg} arise. They are partly inspired by generalized Scherk-Schwarz compactifications which give rise to gauged supergravities not accessible by flux compactifications from the SUGRA regime \cite{Dabholkar:2002sy,Hull:2005hk,Aldazabal:2011nj,Grana:2012rr,Aldazabal:2013mya,Berman:2013cli,Hassler:2014sba}. Some of these backgrounds have an uplift to string theory in terms of left-right asymmetric orbifold constructions \cite{Dabholkar:2002sy,Condeescu:2012sp,Condeescu:2013yma,Hassler:2014sba}, but in general their fate is unknown.

In order to improve this situation three of the authors proposed an alternative theory with a doubled coordinate space called DFT${}_\mathrm{WZW}$ \cite{Blumenhagen:2014gva}. It originates from tree-level Closed String Field Theory (CSFT) calculations up to cubic order in the fields and leading order of $\alpha'$ on a group manifold\footnote{Previous works on duality manifest actions on group manifolds include \cite{Sfetsos:2009vt}.}. This theory is governed by a Wess-Zumino-Witten model on the worldsheet. In DFT${}_\mathrm{WZW}$ the doubling of the coordinates basically refers to the left- and right-moving currents of the WZW model on a group manifold. Interestingly, it turned out that this theory does not reproduce all results known from original DFT: The strong constraint, the gauge transformations and the action receive corrections from the non-trivial string background. Furthermore, the closure of the gauge algebra only requires the strong constraint for fluctuations, whereas the weaker closure constraint is sufficient for the background fields. In this way, one can obtain a consistent tree-level description of non-geometric backgrounds. All these properties suggest that DFT${}_\mathrm{WZW}$ should be considered as a generalization of original DFT. However, a direct comparison between the two at cubic level seems to be impossible. Therefore, in this paper we derive the full generalized metric formulation of the theory. Let us summarize our results  in the following. 

The resulting action to all orders in the fields reads
\begin{equation}\label{eqn:actiongenricciintro}
  S = \int d X^{2D} e^{-2 d} \mathcal{R}\,,
\end{equation}
where $d$ denotes the generalized dilaton and $\mathcal{R}$ represents the generalized curvature scalar
\begin{equation}\label{eqn:gencurvaturecurved}
  \begin{gathered}
    \mathcal{R} = 4 \mathcal{H}^{IJ} \nabla_I \nabla_J d - \nabla_I \nabla_J \mathcal{H}^{IJ} - 4 \mathcal{H}^{IJ} \nabla_I d \nabla_J d + 4 \nabla_I d \nabla_J \mathcal{H}^{IJ} \\ + \frac{1}{8} \mathcal{H}^{KL} \nabla_K \mathcal{H}_{IJ} \nabla_L \mathcal{H}^{IJ} - \frac{1}{2} \mathcal{H}^{IJ} \nabla_J \mathcal{H}^{KL} \nabla_L \mathcal{H}_{IK} +\frac{1}{6} F_{IKL} F_J{}^{KL} \mathcal{H}^{IJ}
  \end{gathered}
\end{equation}
of DFT${}_\mathrm{WZW}$. It incorporates the generalized metric $\mathcal{H}^{IJ}$, the covariant derivative
\begin{equation}\label{eqn:nablaIVJ}
  \nabla_I V^J = \partial_I V^J + \Gamma_{IK}{}^J V^K
\end{equation}
and the structure coefficients $F_{IJK}$ of the group manifold. Both the connection appearing in the covariant derivative and the structure coefficients are determined entirely by the background. In this sense, the theory presented here is manifestly background dependent. As we will discuss in section \ref{sec:DFTWZWtoDFT}, this is not a contradiction in being a generalization of DFT which is background independent once the strong constraint is invoked. We show that the proposed action \eqref{eqn:actiongenricciintro} is invariant under the generalized diffeomorphisms
\begin{align}
  \delta_\xi \mathcal{H}^{IJ} = \mathcal{L}_\xi \mathcal{H}^{IJ} &= 
  \lambda^K \nabla_K \mathcal{H}^{IJ} +  (\nabla^I \lambda_K - \nabla_K \lambda^I) \mathcal{H}^{KJ} + (\nabla^J \lambda_K - \nabla_K \lambda^J) \mathcal{H}^{IK} \nonumber \\
  \delta_\xi d = \mathcal{L}_\xi d &= \xi^I \nabla_I d - \frac{1}{2} \nabla_I \xi^I\,,
\end{align}
where $\mathcal{L}_\xi$ denotes the generalized Lie derivative of the theory. In all calculations, we assume the strong constraint
\begin{equation}
  \nabla_I \partial^I \cdot = 0
\end{equation}
to be fulfilled for the generalized dilaton $d$, the generalized metric $\mathcal{H}^{AB}$, the parameter $\xi^A$ of the generalized Lie derivative and arbitrary products of them. The strong constraint only applies to quantities \underline{in flat indices}. To switch between curved and flat indices the generalized vielbein $E_A{}^I$ of the background is used. Additionally, we also apply the Jacobi identity
\begin{equation}
  F_{IJ}{}^M F_{MK}{}^L + F_{KI}{}^M F_{MJ}{}^L + F_{JK}{}^M F_{MI}{}^L = 0
\end{equation}
for the structure coefficients of the background. Besides generalized diffeomorphisms, \eqref{eqn:actiongenricciintro} is manifestly invariant under 2D-diffeomorphisms
\begin{align}
  \delta_\xi E_A{}^I &= L_\xi E_A{}^I = \xi^J \partial_J E_A{}^I - E_A{}^J \partial_J \xi^I \,,\\
   \delta_\xi e^{-2 d} &= L_\xi e^{-2 d} = \xi^P \delta_P e^{-2 d} + e^{-2 d} \partial_I \xi^I\,,
\end{align}
with $L_\xi$ denoting the ordinary Lie derivative. In view of this, DFT${}_\mathrm{WZW}$ seems to implement a non-trivial extension of the DFT gauge algebra as proposed by Cederwall \cite{Cederwall:2014kxa,Cederwall:2014opa}. Still, there exists an important difference. Whereas Cederwall considered only torsionless covariant derivatives, the covariant derivative \eqref{eqn:nablaIVJ} exhibits a torsionful connection. 

One of the objectives  of this paper is to clarify the relation between  background dependent DFT${}_\mathrm{WZW}$ and  original DFT. We will succeed  to identify DFT${}_\mathrm{WZW}$ with the generalized metric formulation of DFT \cite{Hohm:2010pp} under two special assumptions: First, a distinguished generalized vielbein which fulfills the strong constraint of DFT is required and second an  {\it extended strong constraint}
\begin{equation}
  \partial_I b\, \partial^I f = 0\,,
\end{equation}
linking background fields $b$ and fluctuations $f$, has to be imposed. It is important to note that this constraint is totally optional in the framework of our theory. Hence, it is reasonable to suspect that there exist  valid field configurations in DFT${}_\mathrm{WZW}$ that go beyond DFT. This statement even holds, if the background group manifold is purely geometric or T-dual to a geometric one. Identifying the two theories under the assumptions mentioned above, we confirm the background independence of DFT suggested in \cite{Hohm:2010jy}. This background independence is a result of the very restrictive strong constraint in DFT which renders it equivalent to SUGRA.

The organization of this paper follows the outline given in the last paragraph. After a short review of the DFT${}_\mathrm{WZW}$ cubic action and the required notation, section \ref{sec:genmetricformulation} presents the generalized metric formulation of the action and its gauge transformations. Section \ref{sec:eom} discusses the equations of motion of this action. Further, it derives the generalized curvature scalar and the generalized Ricci tensor. In section \ref{sec:symmetries}, we prove the invariance of the action under generalized diffeomorphisms and 2D-diffeomorphisms. At last, we show the equivalence of our theory and original DFT in section \ref{sec:DFTWZWtoDFT}. A small outlook, discussing the potential and possible applications of DFT${}_\mathrm{WZW}$ concludes the paper in section \ref{sec:outlook}.

\section{Generalized metric formulation}\label{sec:genmetricformulation}
Starting from the results derived in \cite{Blumenhagen:2014gva}, we derive the generalized metric formulation of the DFT${}_\mathrm{WZW}$ action in this section. As a preliminary, subsection \ref{sec:cubicaction} reviews the most important aspects of the cubic action derived in \cite{Blumenhagen:2014gva} and introduces the required notation. Although already discussed in \cite{Blumenhagen:2014gva}, we shortly present the gauge transformations and the C-bracket governing the gauge algebra in subsection \ref{sec:gaugetrafogenmetric} before discussing the new results for action in subsection \ref{sec:actiongenmetric}.

\subsection{Review of cubic action and notation}\label{sec:cubicaction}
The cubic action and gauge transformations were derived at the leading order of $\alpha'$ from CSFT in \cite{Blumenhagen:2014gva}. The starting point are fields $\epsilon^{a\bar b}$ that can be considered as fluctuations around the WZW background. The indices $a $ and $\bar b$ refer to the adjoint representation of the corresponding group $G_L\times G_R$. In addition we also introduce gauge parameters  $\lambda^a$ and $ \lambda^{\bar a} $. In contrast to the toroidal case, one does not consider momentum and winding modes but one considers different representations $\underline R=(\underline r_L,\underline r_R)$ of $G_L\times G_R$. Here, we do not use the form stated in \cite{Blumenhagen:2014gva}, but instead perform the field redefinition
\begin{equation}
  \epsilon^{a\bar b} \rightarrow -2 \epsilon^{a\bar b} \,, \quad
  \lambda^a \rightarrow 2 \lambda^a \quad \text{and} \quad
  \lambda^{\bar a} \rightarrow 2 \lambda^{\bar a} 
\end{equation}
giving rise to
\begin{equation}
  \label{eqn:action}
  \begin{aligned}
  (2 \kappa^2)S &= \int d^{2D} X\sqrt{|H|}\, \Big[ 
      \epsilon_{a\bar b} \,\square \epsilon^{a\bar b} + (D^{\bar b} \epsilon_{a\bar b})^2
      +  (D^a \epsilon_{a \bar b})^2 + 4 \tilde d\, D^a D^{\bar b} \epsilon_{a\bar b} 
      - 4 \tilde d\, \square \tilde d \\[0.1cm]
    & -2 \epsilon_{a\bar b} \bigl( D^a \epsilon_{c\bar d}\, D^{\bar b} \epsilon^{c\bar d} - 
      D^a \epsilon_{c \bar d}\, D^{\bar d} \epsilon^{c\bar b} - D^c \epsilon^{a\bar d}\, D^{\bar b} 
      \epsilon_{c\bar d} \bigr) \\[0.1cm]
    & +2 \epsilon_{a\bar b} \bigl( F^{ac}{}_d\, D^{\bar e} \epsilon^{d\bar b} \;\epsilon_{c\bar e} 
      + F^{\bar b\bar c}{}_{\bar d}\, D^e \epsilon^{a\bar d}\; \epsilon_{e\bar c} \bigr)
      + \frac{2}{3} F^{ace}\, F^{\bar b\bar d\bar f}\, \epsilon_{a\bar b}\, \epsilon_{c\bar d}\,
      \epsilon_{e\bar f} \\[0.1cm]
    & +\tilde d \bigl( 2 (D^a \epsilon_{a\bar b})^2 + 2 (D^{\bar b} \epsilon_{a\bar b})^2 + 
      (D_c \epsilon_{a\bar b})^2 + (D_{\bar c} \epsilon_{a\bar b})^2 + 4 \epsilon^{a\bar b}
      ( D_a D^c \epsilon_{c\bar b} + D_{\bar b} D^{\bar c} \epsilon_{a\bar c} )
      \bigr) \\[0.1cm]
    & -8 \epsilon_{a\bar b}\, \tilde d\, D^a D^{\bar b} \tilde d + 4 {\tilde d}^2\, \square \tilde d \Big]
  \end{aligned}
\end{equation}
with the abbreviation
\begin{equation}
  \Box = \frac{1}{2} ( D_a D^a + D_{\bar a} D^{\bar a} )
\end{equation}
and the corresponding gauge transformations
\begin{equation}
  \label{eqn:gaugetrafo}
  \begin{aligned}
  \delta_{\lambda} \epsilon^{a\bar b} =& - D^{\bar b} \lambda^a + 
    D^a \lambda_c \epsilon^{c\bar b} - D_c \lambda^a \epsilon^{c\bar b} + \lambda^c D_c \epsilon^{a \bar b}
    + F^a{}_{cd}\, \lambda^c \epsilon^{d\bar b}  + \\ 
  & -D^a \lambda^{\bar b} + D^{\bar b} \lambda_{\bar c} \epsilon^{a\bar c} - 
    D_{\bar c} \lambda^{\bar b} \epsilon^{a \bar c} + \lambda^{\bar c} D_{\bar c} \epsilon^{a \bar b}
    + F^{\bar b}{}_{\bar c \bar d} \lambda^{\bar c}\, \epsilon^{a \bar d} \,,\\
    \delta_\lambda \tilde d =& -\frac{1}{2} D_a \lambda^a + \lambda_a\, D^a \tilde d - 
    \frac{1}{2} D_{\bar a} \lambda^{\bar a} + \lambda_{\bar a}\,
    D^{\bar a} {\tilde d} \,.
  \end{aligned}
\end{equation}
Besides, further rescaling in the definitions given later in this subsection, this field redefinition helps to get rid of a $1/2$ factor which arises in \cite{Blumenhagen:2014gva} between the DFT and the DFT${}_\mathrm{WZW}$ results. To allow a clear distinction between background fields and fluctuations, we have changed the notation of \cite{Blumenhagen:2014gva}. Now, $\tilde d$ denotes fluctuations of the generalized dilaton $d = \bar d + \tilde d$ which combines the background field $\bar d$ and the fluctuations. As a consequence of level-matching in closed string theory, the fields $\epsilon_{a\bar b}$, $\tilde d$ and the gauge parameters $\lambda_a$ and $\lambda_{\bar a}$ have to fulfill the strong constraint
\begin{equation}\label{eqn:strongconstcomp}
  ( D_a D^a - D_{\bar a} D^{\bar a} ) \cdot = 0 \,,
\end{equation}
where $\cdot$ not only denotes the mentioned field but also arbitrary products of them.

On the world sheet, the theory is governed by a CFT with two independent, a chiral (left mover) and an anti-chiral (right mover), Ka\v{c}-Moody current algebras. The structure coefficients of their central extensions $\mathfrak{g}_{\rm L}\times \mathfrak{g}_{\rm R}$ are denoted by $F_{ab}{}^c$ and $F_{\bar a\bar b}{}^{\bar c}$. Bared and unbared indices allow to distinguish between the algebras for the left and right moving part of the closed string. These indices run from $1\dots D$, the dimension of the group manifold used as background. The integration in \eqref{eqn:action} is performed over a product manifold combining the Lie groups $G_{\rm L}\times G_{\rm R}$ associated to the Lie algebras $\mathfrak{g}_{\rm L}\times \mathfrak{g}_{\rm R}$. This manifold is parameterized by the $2D$ coordinates $X^I=( x^i \,\,\, x^{\bar i})$ and is equipped with the metric
\begin{equation}\label{eqn:SAB}
  S_{AB} = 2 \begin{pmatrix}
    \eta_{ab} & 0 \\
    0 & \eta_{\bar a\bar b}
  \end{pmatrix} \quad \text{and its inverse} \quad
  S^{AB} = \frac{1}{2} \begin{pmatrix}
    \eta^{ab} & 0 \\
    0 & \eta^{\bar a\bar b}
  \end{pmatrix}
\end{equation}
in flat indices. It combines the killing metrics $\eta_{ab}$ / $\eta_{\bar a\bar b}$ of the Lie algebras $\mathfrak{g}_{\rm L}$ / $\mathfrak{g}_{\rm R}$ which are used to lower flat indices. Moreover, it is very convenient to introduce the vielbein
\begin{equation}\label{eqn:EAI}
  E_A{}^I = \begin{pmatrix}
    e_a{}^i & 0 \\
    0 & e_{\bar a}{}^{\bar i}
  \end{pmatrix} \quad \text{and its inverse transposed} \quad
  E^A{}_I = \begin{pmatrix}
    e^a{}_i & 0 \\
    0 & e^{\bar a}{}_{\bar i}
  \end{pmatrix} 
\end{equation}
in order to switch between flat and curved indices. Applying it on the partial derivatives $\partial_I=(\partial_i \,\,\, \partial_{\bar i})$ of the background manifold $G_{\rm L}\times G_{\rm R}$, it gives rise to the doubled flat derivative 
\begin{equation}\label{eqn:DA}
  D_A= E_A{}^I \partial_I = (D_a \,\,\, D_{\bar a})\,.
\end{equation}
Finally $H_{IJ}$, whose determinante $H$ is used in \eqref{eqn:action}, is defined as the curved version
\begin{equation}
  H_{IJ}=E^A{}_I S_{AB} E^B{}_J
\end{equation}
of $S_{AB}$. As a consequence of the rescaled flat metric $S_{AB}$, $H_{IJ}$ differs by a factor 2 from the definition in \cite{Blumenhagen:2014gva}. To keep the action integral \eqref{eqn:action} invariant, one has to perform the compensating change of variables $X^I \rightarrow X^I / \sqrt{2}$.  Besides the background metric $S_{AB}$ in flat indices, it is convenient to introduce the metric
\begin{equation}\label{eqn:etaAB}
  \eta_{AB} = 2 \begin{pmatrix}
    \eta_{ab} & 0 \\
    0 & -\eta_{\bar a\bar b}
  \end{pmatrix} \quad \text{and its inverse} \quad
  \eta^{AB} = \frac{1}{2} \begin{pmatrix}
    \eta^{ab} & 0 \\
    0 & -\eta^{\bar a\bar b}
  \end{pmatrix}
\end{equation}
to lower and raise doubled indices. In combination with the doubled flat derivative \eqref{eqn:DA}, it e.g. allows to express the strong constraint \eqref{eqn:strongconstcomp} in the compact form
\begin{equation}\label{eqn:strongconst}
  \eta_{AB} D_A D_B\, \cdot = D_A D^A\, \cdot = 0\,.
\end{equation}

\subsection{Gauge transformations}\label{sec:gaugetrafogenmetric}
Switching from the notation with bared and unbared indices to doubled indices, generally simplifies the equations in DFT${}_\mathrm{WZW}$ a lot. In this respect, the strong constraint \eqref{eqn:strongconst} is a toy example. More drastic is the effect on the gauge transformations \eqref{eqn:gaugetrafo}. In order to express them in doubled notation, we follow \cite{Blumenhagen:2014gva} and introduce the symmetric, $O(D,D)$ matrix 
\begin{equation}\label{eqn:HAB}
  \mathcal{H}^{AB} = \text{exp}(\epsilon^{AB})
    = S^{AB} + \epsilon^{AB} + \frac{1}{2} \epsilon^{AC} S_{CD} \epsilon^{DB} + \frac{1}{6} \epsilon^{AC} S_{CD} \epsilon^{DE} S_{EF} \epsilon^{FB} + \dots\,,
\end{equation}
called generalized metric. It is generated by 
\begin{equation}
\epsilon^{AB} = \begin{pmatrix}
0 & \epsilon^{a\bar{b}} \\ \epsilon^{\bar{a}b} & 0 
\end{pmatrix} \quad \text{with} \quad \epsilon^{a\bar{b}} = ({\epsilon^T})^{\bar{b}a}\,,
\end{equation}
which embeds the fluctuations $\epsilon^{a\bar b}$ into a tensor with doubled indices. Furthermore, we define the flat covariant derivatives 
\begin{equation}\label{eqn:flatcovderv}
  \nabla_A V^B = D_A V^B + \frac{1}{3} F^B{}_{AC} V^C \quad \text{and} \quad
  \nabla_A V_B = D_A V_B + \frac{1}{3} F_{BA}{}^C V_C \,,
\end{equation}
where
\begin{equation}\label{eqn:FABC}
  F_{AB}{}^C = \begin{cases}
    F_{ab}{}^c \\ F_{\bar a\bar b}{}^{\bar c} \\ 0 \quad \quad \text{otherwise}
  \end{cases}
\end{equation}
combines the structure coefficients defining the Ka\v{c}-Moody algebras for the strings left and right moving parts. At this point, let us recall  the conventions from \cite{Blumenhagen:2014gva}: $D_A$, $F_{AB}{}^C$ and $\xi^A$ are considered as ``fundamental'' objects, meaning their bared and unbared components do not receive additional minus signs or prefactors. From these quantities all others are derived by raising/lowering the doubled indices with the $\eta$-metric. A simple example is
\begin{equation}
  \xi^A = ( \xi^a \,\,\, \xi^{\bar a} ) \quad \text{and} \quad
  \xi_A = \xi^B \eta_{BA} = ( 2 \xi_a \,\,\, -2 \xi_{\bar a} )\,.
\end{equation}
Now, we expand the generalized metric \eqref{eqn:HAB} into components
\begin{equation}\label{eqn:HABexpansion}
\mathcal{H}^{AB} = \begin{pmatrix}
  \frac{1}{2}\eta^{ab} + \epsilon^{a\bar{c}} \eta_{\bar{c}\bar{d}} \epsilon^{\bar{d}b} & \epsilon^{a\bar{b}} + \frac{2}{3} \epsilon^{a\bar{c}} \eta_{\bar{c}\bar{d}} \epsilon^{\bar{d}e} \eta_{ef} \epsilon^{f\bar{b}} \\ \epsilon^{\bar{a}b} + \frac{2}{3} \epsilon^{\bar{a}c} \eta_{cd} \epsilon^{d\bar{e}} \eta_{\bar{e}\bar{f}} \epsilon^{\bar{f}b} & \frac{1}{2} \eta^{\bar{a}\bar{b}} + \epsilon^{\bar{a}c} \eta_{cd} \epsilon^{d\bar{b}}
\end{pmatrix} + \mathcal{O}(\epsilon^4)
\end{equation}
up to cubic order in the fields. Plugging this expansion into
\begin{align}\label{eqn:gendiffHAB&tilded}
  \delta_\xi \mathcal{H}^{AB} = \mathcal{L}_\xi \mathcal{H}^{AB} &= \lambda^C \nabla_C \mathcal{H}^{AB} + (\nabla^A \lambda_C - \nabla_C \lambda^A) \mathcal{H}^{CB} + (\nabla^B \lambda_C - \nabla_C \lambda^B) \mathcal{H}^{AC} \nonumber \\
  \delta_\xi \tilde d = \mathcal{L}_\xi \tilde d & = \xi^A D_A \tilde d - \frac{1}{2} D_A \xi^A \,,
\end{align}
one recovers the gauge transformations \eqref{eqn:gaugetrafo} up to additional terms which are not linear in the field or the gauge parameter. The same holds for the C-bracket
\begin{equation}
  [\xi_1, \xi_2]_C^A = \xi_1^B \,\nabla_B \xi_2^A - \frac{1}{2} \xi_1^B\, \nabla^A\, \xi_{2\,B} - (1 \leftrightarrow 2)\,.
\end{equation}

\subsection{Action}\label{sec:actiongenmetric}
In this subsection, we rewrite the action \eqref{eqn:action} in terms of the generalized metric. The guiding principle is inspired by the results for the gauge transformations and the C-bracket discussed in the last subsection: In the expressions known from traditional DFT, one has to substitute partial derivatives by covariant derivatives \eqref{eqn:flatcovderv}. Taking into account the original DFT action in the generalized metric formulation \cite{Hohm:2010pp} and following this principle, the action should read
\begin{align}
  S = \int d^{2n} X e^{-2d} \Big( & \frac{1}{8} \mathcal{H}^{CD} \nabla_C \mathcal{H}_{AB} \nabla_D \mathcal{H}^{AB} -\frac{1}{2} \mathcal{H}^{AB} \nabla_{B} \mathcal{H}^{CD} \nabla_D \mathcal{H}_{AC} \nonumber \\
  \label{eqn:actiongenmetricnaive}
  & - 2 \nabla_A d \nabla_B \mathcal{H}^{AB} + 4 \mathcal{H}^{AB} \nabla_A d \nabla_B d \Big)\,.
\end{align}

Subsequently, we prove that, up to cubic terms in the fields, this action indeed reproduces \eqref{eqn:action} up to a missing term that has to be added to \eqref{eqn:actiongenmetricnaive}. To keep this straightforward though  cumbersome calculation as traceable as possible, we begin with terms containing two flat derivatives like e.g.
\begin{equation}\label{eqn:HDHDH}
  e^{-2 d} \frac{1}{8} \mathcal{H}^{CD} D_C \mathcal{H}_{AB} D_D \mathcal{H}^{AB}.
\end{equation}
We further simplify the calculation by first considering the term
\begin{equation}
  \frac{1}{8} S^{CD} D_C \mathcal{H}_{AB} D_D \mathcal{H}^{AB},
\end{equation}
which gives rise to
\begin{align} \label{eqn:SDHDH1}
  \frac{1}{8} S^{CD} D_C \mathcal{H}_{AB} D_D \mathcal{H}^{AB} &= -\frac{1}{2} \Big( D_c \epsilon_{a\bar{b}} D^c \epsilon^{a\bar{b}} + D_{\bar{c}} \epsilon_{a\bar{b}} D^{\bar{c}} \epsilon^{a\bar{b}} \Big) + \mathcal{O}(\epsilon^4) \nonumber \\
  &= \epsilon_{a\bar{b}}\Box \epsilon^{a\bar{b}} - \epsilon_{a\bar{b}} D_c \tilde{d} D^c \epsilon^{a\bar{b}} - \epsilon_{a\bar{b}} D_{\bar{c}} \tilde{d} D^{\bar{c}} \epsilon^{a\bar{b}} + \mathcal{O}(\epsilon^4)\,,
\end{align}
after plugging in the components of $S^{AB}$ and $\mathcal{H}^{AB}$, according to \eqref{eqn:SAB} and \eqref{eqn:HABexpansion}. From the first to the second line in \eqref{eqn:SDHDH1}, we perform integration by parts by applying the rule
\begin{equation}\label{eqn:ibp}
  \int d^{2n} X e^{-2d} u D_a v = - \int d^{2n} \sqrt{|H|} e^{-2 \tilde d} ( -2 u D_a \tilde d  + D_a u ) v = \int d^{2n} X e^{-2d} (2 u D_a \tilde d   - D_a u) v \,.
\end{equation}
It automatically arises, if one splits the generalized dilaton
\begin{equation}\label{eqn:splitdilaton}
  d = \bar d + \tilde d = -\frac{1}{2} \log \sqrt{|H|} + \tilde d
\end{equation}
into the background part $\bar d$ and the fluctuations $\tilde d$ around this background. Performing integrations by parts again and dropping the terms in quartic order in the fields, we obtain
\begin{equation}\label{eqn:SDHDH2}
  \frac{1}{8} S^{CD} D_C \mathcal{H}_{AB} D_D \mathcal{H}^{AB} = \epsilon_{a\bar{b}} \Box \epsilon^{a\bar{b}} + \tilde{d} \big( D_c \epsilon_{a\bar{b}} \big)^2 + \tilde{d} \big( D_{\bar{c}} \epsilon_{a\bar{b}} \big)^2 + 2 \tilde{d}\, \epsilon_{a\bar{b}} \Box \epsilon^{a\bar{b}}  + \mathcal{O}(\epsilon^4) + \mathcal{O}(\tilde d^2 \epsilon^2)\,.
\end{equation}
Now, it is straightforward to read off the remaining terms of \eqref{eqn:HDHDH}, namely
\begin{equation}
  e^{-2 d} \frac{1}{8} \mathcal{H}^{CD} D_C \mathcal{H}_{AB} D_D \mathcal{H}^{AB} = \sqrt{H} \Big[ \epsilon_{a\bar{b}} \Box \epsilon^{a\bar{b}} -2 \epsilon^{c\bar{d}} D_c \epsilon_{a\bar{b}} D_{\bar{d}} \epsilon^{a\bar{b}} + \tilde{d} \big( D_c \epsilon_{a\bar{b}} \big)^2 + \tilde{d} \big( D_{\bar{c}} \epsilon_{a\bar{b}} \big)^2 \Big]\,.
\end{equation}
Here and in the following, $\mathcal{O}(\dots)$ is suppressed for brevity. The last term in \eqref{eqn:SDHDH2} cancels against  a term arising in the expansion of
\begin{equation}
  e^{-2 d} = \sqrt{|H|} \big(1 - 2 \tilde d + 2 \tilde d^2 + \dots \big)\,.
\end{equation}
Next, we turn to the term
\begin{equation}
  -\frac{1}{2} \mathcal{H}^{AB} D_{B} \mathcal{H}^{CD} D_D \mathcal{H}_{AC}
\end{equation}
for which the calculations are more cumbersome. Using the commutation relations for flat derivatives
\begin{equation}
  [ D_a, D_b ] = F_{ab}{}^c D_c\, ,\quad [ D_{\bar a}, D_{\bar b} ] = F_{\bar{a}\bar{b}}{}^{\bar c} D_{\bar c}
\end{equation}
and performing integration by parts, we finally obtain the result
\begin{align}
  - e^{-2 d} \frac{1}{2} \mathcal{H}^{AB} D_{B} \mathcal{H}^{CD} D_D \mathcal{H}_{AC} &= \sqrt{|H|} \Big[ \big( D^a e_{a\bar{b}} \big)^2 + \big( D^{\bar{b}} e_{a\bar{b}} \big)^2 \nonumber \\ 
  &\,- \big( F_d{}^{ac} D_c \epsilon^{d\bar{b}}  \epsilon_{a\bar{b}}\,+ F_{\bar d}{}^{\bar a\bar c} D_{\bar{c}} \epsilon^{b\bar{d}} \epsilon_{b\bar{a}} \big) \big( 1-2\tilde{d} \big) \nonumber \\ 
  &\,+ 2 \tilde{d}\, \epsilon^{a\bar{b}} D_a D^c \epsilon_{c\bar{b}} - 2 \tilde{d} D_a D^c \epsilon^{a\bar{b}} \epsilon_{c\bar{b}} - 2 \tilde{d} D^c \epsilon^{a\bar{b}}  D_a \epsilon_{c\bar{b}} \nonumber \\ 
  &\,+2 \tilde d\, \epsilon^{a\bar{b}} D_{\bar{b}} D^{\bar{c}} \epsilon_{a\bar{c}} - 2 \tilde{d} D_{\bar{b}} D^{\bar{c}} \epsilon^{a\bar{b}} \epsilon_{a\bar{c}} - 2 \tilde{d} D^{\bar{c}} \epsilon^{a\bar{b}} D_{\bar{b}} \epsilon_{a\bar{c}} \nonumber \\ 
  &\,+ 2 \epsilon_{a\bar{b}} \big( D^a \epsilon_{c\bar{d}} D^{\bar{d}} \epsilon^{c\bar{b}} + D^c \epsilon^{a\bar{d}} D^{\bar{b}} \epsilon_{c\bar{d}} \big) \Big].
\end{align}
All remaining terms in the action \eqref{eqn:actiongenmetricnaive} contain covariant derivatives acting on the generalized dilaton $d$. Its background part $\bar d$ is covariantly constant and the fluctuations $\tilde d$ transform like a scalar. Thus, we are able to identify
\begin{equation} \label{eqn:nablaAd}
  \nabla_A d = D_A \tilde d\,.
\end{equation}
In combination with the expansion \eqref{eqn:HABexpansion} of $\mathcal{H}^{AB}$, this identity gives rise to
\begin{equation}
  4 \mathcal{H}^{AB} D_A \tilde d D_B \tilde d = 2 D_a \tilde d D^a \tilde d +  2 D_{\bar a} \tilde d D^{\bar a} \tilde d + 8 \epsilon^{a\bar b} D_a \tilde d D_{\bar b} \tilde d\,.
\end{equation}
Taking into account the prefactor $e^{-2d}$, we obtain
\begin{equation}
  e^{-2 d} 4 \mathcal{H}^{AB} D_A \tilde{d} D_B \tilde{d} = \sqrt{|H|} \big[ -4\tilde{d} \Box \tilde{d} + 8 \epsilon^{a\bar{b}} D_a \tilde{d} D_{\bar{b}} \tilde{d}  + 4 \tilde{d}^2 \Box \tilde{d} \big]\,,
\end{equation}
where we applied the relation
\begin{equation}
  \sqrt{H} 4 \tilde{d}^2 \Box d = \sqrt{H} \big(-4\tilde{d} D_a \tilde{d} D^a \tilde{d} - 4\tilde{d} D_{\bar{a}} \tilde{d} D^{\bar{a}} \tilde{d}\;\big)\,.
\end{equation}
The last term in \eqref{eqn:actiongenmetricnaive}, which contains two flat derivatives, gives rise to
\begin{align}
  - e^{-2d} 2 D_A \tilde{d} D_B \mathcal{H}^{AB} &= \sqrt{|H|} \Big[ 4 \tilde{d} D_a D_{\bar{b}} \epsilon^{a\bar{b}} - 8\epsilon^{a\bar{b}} D_a \tilde{d} D_{\bar{b}} \tilde{d} - 8\tilde{d} \epsilon^{a\bar{b}} D_a D_{\bar{b}} \tilde{d} \nonumber \\ 
  &\,+ 2 \tilde{d} \big( D^a \epsilon_{a\bar{b}} \big)^2 + 2 \tilde{d} \big( D^{\bar{b}} \epsilon_{a\bar{b}} \big)^2 + 2 \tilde{d} \epsilon^{a\bar{b}} D_a D^c \epsilon_{a\bar{b}} + 2 \tilde{d} \epsilon^{a\bar{b}} D_{\bar{b}} D^{\bar{c}} \epsilon_{a\bar{c}} \nonumber \\ 
  &\,+2 \tilde{d} D^c \epsilon^{a\bar{b}} D_a \epsilon_{c\bar{b}} + 2 \tilde{d} D^{\bar{c}} \epsilon^{a\bar{b}}  D_{\bar{b}} \epsilon_{a\bar{c}} + 2 \tilde{d}  D_a D^c \epsilon^{a\bar{b}} \epsilon_{c\bar{b}} \nonumber \\
  &\,+ 2 \tilde{d} \big( D_{\bar{b}} D^{\bar{c}} \epsilon^{a\bar{b}} \big) \epsilon_{a\bar{c}}\Big]\,.
\end{align}
Now, we are done with all terms required for the abelian case $F_{ABC}=0$. Hence, it is a convenient check of the results obtained so far to write down the complete abelian action
\begin{align}
  \left. S\right|_{F_{ABC}=0} &= \int d^{2n}X \sqrt{|H|} \Big[ \epsilon_{a\bar{b}} \Box \epsilon^{a\bar{b}} +  \big( D^a e_{a\bar{b}} \big)^2 + \big( D^{\bar{b}} e_{a\bar{b}} \big)^2  + 4\tilde{d} D_a D_{\bar{b}} \epsilon^{a\bar{b}} -4\tilde{d} \Box \tilde{d} \nonumber \\ & -2 \epsilon_{a\bar{b}} \big( D_a \epsilon_{c\bar{d}} D_{\bar{b}} \epsilon^{c\bar{d}} - D^a \epsilon_{c\bar{d}} D^{\bar{d}} \epsilon^{c\bar{b}} - D^c \epsilon^{a\bar{d}} D^{\bar{b}} \epsilon_{c\bar{d}} \big) \nonumber \\ & + \tilde{d} \Big( 2 \big( D^a e_{a\bar{b}} \big)^2  + 2 \big( D^{\bar{b}} e_{a\bar{b}} \big)^2 + \big( D_c \epsilon_{a\bar{b}} \big)^2 + 
  \big( D_{\bar{c}} \epsilon_{a\bar{b}} \big)^2 + 4 \epsilon^{a\bar{b}} \big( D_a D^c \epsilon_{a\bar{b}} + D_{\bar{b}} D^{\bar{c}} \epsilon_{a\bar{c}} \big) \Big) \nonumber \\ & + 4\tilde{d}^2 \Box \tilde{d} - 8\tilde{d} \epsilon^{a\bar{b}} D_a D_{\bar{b}} \tilde{d} \Big]\,.
\end{align}
It indeed matches with the action \eqref{eqn:action} after dropping all terms depending on the structure coefficients $F_{abc}$ and $F_{\bar a\bar b\bar c}$. 

Let us now consider these terms so that  we have to consider the full covariant derivative instead of only using its flat derivative part. Let us start with
\begin{align}
- 2 \nabla_A d \nabla_B \mathcal{H}^{AB} &= - D_A \tilde{d} \Big( D_B \mathcal{H}^{AB} + \frac{1}{3} \big( {F^A}_{BC} \mathcal{H}^{CB} + {F^B}_{BC} \mathcal{H}^{AC} \big) \Big) \nonumber \\ 
  &= - D_A \tilde{d} D_B \mathcal{H}^{AB}\,,
\end{align}
where the second term in the first line vanishes due the total antisymmetry of $F_{ABC}$ and the symmetry of $\mathcal{H}^{AB}$. The third term is zero due to the unimodularity condition
\begin{equation}
  F^A{}_{AB} = 0\,,
\end{equation}
which the structure coefficients have to fulfill \cite{Blumenhagen:2014gva}. At this point, we come to the more challenging part
\begin{equation}\label{eqn:Htermscov}
  \frac{1}{8} \mathcal{H}^{CD} \nabla_C \mathcal{H}_{AB} \nabla_D \mathcal{H}^{AB} -\frac{1}{2} \mathcal{H}^{AB} \nabla_{B} \mathcal{H}^{CD} \nabla_D \mathcal{H}_{AC}\,.
\end{equation}
In the subsequent computation, we ignore all terms which contain more than one flat derivative, because we already discussed these contributions above. The first part of \eqref{eqn:Htermscov} gives rise to
\begin{align}
  \frac{1}{8} \mathcal{H}^{CD} \nabla_C & \mathcal{H}_{AB} \nabla_D \mathcal{H}^{AB} = \frac{1}{12} \mathcal{H}^{CD} D_C \mathcal{H}_{AB} {F^{A}}_{DE} \mathcal{H}^{EB} + \frac{1}{12} \mathcal{H}^{CD} {F_{AC}}^{F} \mathcal{H}_{FB} D_D \mathcal{H}^{AB} \nonumber \\ & +\frac{1}{36} \mathcal{H}^{CD} {F_{AC}}^{F} \mathcal{H}_{FB} {F^{A}}_{DE} \mathcal{H}^{EB} + \frac{1}{36} \mathcal{H}^{CD} {F_{AC}}^{F} \mathcal{H}_{FB} {F^{B}}_{DE} \mathcal{H}^{AE} + \mathcal{O}(D^2), \label{eqn:HCHCH}
\end{align}
where the second term on the right hand side is equivalent to
\begin{equation}
  \mathcal{H}^{CD} {F_{AC}}^{F} \mathcal{H}_{FB} D_D \mathcal{H}^{AB} = \mathcal{H}^{CD} {F^{A}}_{DE} \mathcal{H}^{EB} D_C \mathcal{H}_{AB}\,,
\end{equation}
after using the symmetry of $\mathcal{H}^{AB}$ and relabeling the indices. For the fourth term, we use the
total antisymmetry of the structure coefficients to yield
\begin{equation}
    \mathcal{H}^{CD} {F_{AC}}^{F} \mathcal{H}_{FB} {F^{B}}_{DE} \mathcal{H}^{AE} = - F_{ACE} F_{BDF} \mathcal{H}^{AB} \mathcal{H}^{CD} \mathcal{H}^{EF} \,.
\end{equation}
Applying these two substitutions, \eqref{eqn:HCHCH} simplifies to
\begin{align}
  \frac{1}{8} \mathcal{H}^{CD} \nabla_C \mathcal{H}_{AB} \nabla_D \mathcal{H}^{AB} &\,= \frac{1}{6} \mathcal{H}^{CD} D_C \mathcal{H}_{AB} {F^{A}}_{DE} \mathcal{H}^{EB} +\frac{1}{36} \mathcal{H}^{CD} {F_{AC}}^{F} {F^{A}}_{DE} \mathcal{H}_{FB} \mathcal{H}^{EB} \nonumber \\ &\quad\; - \frac{1}{36} F_{ACE} F_{BDF} \mathcal{H}^{AB} \mathcal{H}^{CD} \mathcal{H}^{EF} + \mathcal{O}(D^2)\,.
\end{align}
For the second part of \eqref{eqn:HCHCH}, we obtain in a similar fashion
\begin{align}
-\frac{1}{2} \mathcal{H}^{AB} \nabla_{B} \mathcal{H}^{CD} \nabla_D \mathcal{H}_{AC} 
&\,= \frac{1}{3} \mathcal{H}^{CD} D_C \mathcal{H}_{AB} {F^{A}}_{DE} \mathcal{H}^{EB} - \frac{1}{6} \mathcal{H}^{AB} D_B \mathcal{H}^{CD} {F_{CD}}^E \mathcal{H}_{AE} \nonumber \\ &\;-\frac{1}{6} \mathcal{H}^{AB} {F^D}_{BE} \mathcal{H}^{CE} D_D \mathcal{H}_{AC} - \frac{1}{18} F_{ACE} F_{BDF} \mathcal{H}^{AB} \mathcal{H}^{CD} \mathcal{H}^{EF}  \nonumber \\ &\;+\frac{1}{18} \mathcal{H}^{CD} {F_{AC}}^{F} {F^{A}}_{DE} \mathcal{H}_{FB} \mathcal{H}^{EB} + \mathcal{O}(D^2)\,.
\end{align}
After combining these results, we finally get
\begin{align}\label{eqn:Ftermsaction}
  e^{-2 d} \Big[ \frac{1}{8} \mathcal{H}^{CD} &\nabla_C \mathcal{H}_{AB} \nabla_D \mathcal{H}^{AB} -\frac{1}{2} \mathcal{H}^{AB} \nabla_{B} \mathcal{H}^{CD} \nabla_D \mathcal{H}_{AC}\Big] \nonumber\\ &\,= \sqrt{|H|} \Big[ 2 \epsilon_{a\bar{b}} \big( {F^{ac}}_d D^{\bar{e}} \epsilon^{d\bar{b}} \epsilon_{c\bar{e}} + {F^{\bar{b}\bar{c}}}_{\bar{d}} D^{e} \epsilon^{a\bar{d}} \epsilon_{e\bar{c}} \big) + \frac{2}{3} F_{ace} F_{\bar{b}\bar{d}\bar{f}} \epsilon^{a\bar{b}} \epsilon^{c\bar{d}} \epsilon^{e\bar{f}} \nonumber \\ &\;-\frac{1}{6} \big( F^a{}_{cd} F_b{}^{cd} \epsilon_{a\bar e} \epsilon^{b \bar e} + F^{\bar a}{}_{\bar c\bar d} F_{\bar b}{}^{\bar c\bar d} \epsilon_{e \bar a} \epsilon^{e \bar b} \big) \big( 1 - 2\tilde{d} \big) + \mathcal{O}(D^2) \Big]\,.
\end{align}
The first line on the right hand side exactly reproduces the structure coefficients dependent terms in the cubic action \eqref{eqn:action}, but the second line has to be canceled to successfully reproduce the action. Achieving this is done by adding the term
\begin{equation}
  \frac{1}{6} F_{ACE} F_{BDF} \mathcal{H}^{AB} \eta^{CD} \eta^{EF} + V_0 =
    \frac{1}{6} \big( F^a{}_{cd} F_b{}^{cd} \epsilon_{a\bar e} \epsilon^{b \bar e} +
      F^{\bar a}{}_{\bar c\bar d} F_{\bar b}{}^{\bar c\bar d} \epsilon_{e \bar a} \epsilon^{e \bar b}
      \big)
\end{equation}
with
\begin{align}
  V_0 &= - \frac{1}{6} F_{ACE} F_{BDF} S^{AB} S^{CD} S^{EF} \nonumber \\
      &= - \frac{1}{4} F_{ACE} F_{BDF} S^{AB} \eta^{CD} \eta^{EF} +
         \frac{1}{12} F_{ACE} F_{BDF} S^{AB} S^{CD} S^{EF}
\end{align}
to the naive action \eqref{eqn:actiongenmetricnaive}. To obtain the second line in the expression for $V_0$, we applied the identity 
\begin{equation}
  F_{ACE} F_{BDF} S^{CD} S^{EF} =  F_{ACE} F_{BDF} \eta^{CD} \eta^{EF}
\end{equation}
which holds due to the strict separation of bared and unbared structure coefficients \eqref{eqn:FABC}. Substituting the structure coefficients $F_{ABC}$ by the covariant fluxes $\mathcal{F}_{ABC}$, $V_0$ matches the vacuum expectation value of the scalar potential arising from generalized Scherk-Schwarz compactifications \cite{Geissbuhler:2011mx,Aldazabal:2011nj}. Note that even though we do not impose the strong constraint on the background, $V_0$ lacks the $1/6 F_{ABC} F^{ABC}$ introduced by hand in the flux formulation \cite{Grana:2012rr,Geissbuhler:2013uka} in order to reproduce the scalar potential of half-maximal, electrically gauge supergravities \cite{Schon:2006kz}. If we consider the full $2D$-dimensional doubled space time instead of only its $2n$-dimensional compact subspace, $V_0$ has to vanish for each background which gives rise to a well defined CFT. Otherwise the combined central charge of the ghost system and the bosons would not vanish.

We close this section with the complete action of DFT${}_\mathrm{WZW}$
\begin{equation}
\begin{aligned}
  S = & \int d^{2D} X e^{-2d} \Big(  \frac{1}{8} \mathcal{H}^{CD} \nabla_C \mathcal{H}_{AB} \nabla_D \mathcal{H}^{AB} -\frac{1}{2} \mathcal{H}^{AB} \nabla_{B} \mathcal{H}^{CD} \nabla_D \mathcal{H}_{AC} \\
  \label{eqn:actiongenmetric}
  & - 2 \nabla_A d \nabla_B \mathcal{H}^{AB} + 4 \mathcal{H}^{AB} \nabla_A d \nabla_B d + \frac{1}{6} F_{ACD} F_B{}^{CD} \mathcal{H}^{AB} \Big)\,,
\end{aligned}
\end{equation}
in the generalized metric formulation. For obtaining the action in curved indices, one has to remember the vielbein compatibility condition $\nabla_I E_A{}^J = 0$ of the covariant derivative. Due to this condition it is legitimate to simply substitute all flat indices with curved ones.

\section{Equations of motion}\label{sec:eom}
After deriving the full action of DFT${}_\mathrm{WZW}$ in the last section, we now discuss its equations of motion. It is convenient to split them into two independent parts. First, we present the variation of the action \eqref{eqn:actiongenmetric} with respect to the generalized dilaton $d$ in subsection \ref{sec:gencurvature}. It gives rise to the generalized curvature scalar $\mathcal{R}$. Furthermore, we show how the action can be rewritten in terms of this scalar. In the second step, we perform the variation with respect to the generalized metric $\mathcal{H}^{AB}$ in subsection \ref{sec:genriccitensor}. Just as in the generalized metric formulation of DFT \cite{Hohm:2010pp}, we have to apply an appropriate projection, taking into account the $O(D,D)$ property of the generalized metric, to obtain the generalized Ricci tensor $\mathcal{R}_{IJ}$.

\subsection{Generalized curvature scalar}\label{sec:gencurvature}
Following \cite{Hohm:2010pp}, we define the generalized scalar curvature $\mathcal{R}$ of DFT${}_\mathrm{WZW}$ using the variation of the action \eqref{eqn:actiongenmetric}
\begin{equation}\label{eqn:variationd}
  \delta S = - 2 \int d^{2D} X\, e^{-2d}\, \mathcal{R}\, \delta d,
\end{equation}
with respect to the generalized dilaton $d$. A straightforward calculation gives rise to
\begin{equation}
  \begin{aligned}
  \label{eqn:gencurvature}
  \mathcal{R} &=  4 \mathcal{H}^{AB} \nabla_A \nabla_B d - \nabla_A \nabla_B \mathcal{H}^{AB} - 4 \mathcal{H}^{AB} \nabla_A d\, \nabla_B d + 4 \nabla_A d \,\nabla_B \mathcal{H}^{AB} \\ &\,+ \frac{1}{8} \mathcal{H}^{CD} \nabla_C \mathcal{H}_{AB} \nabla_D \mathcal{H}^{AB} - \frac{1}{2} \mathcal{H}^{AB} \nabla_B \mathcal{H}^{CD} \nabla_D \mathcal{H}_{AC} +\frac{1}{6} F_{ACD} F_B{}^{CD} \mathcal{H}^{AB}\,.
  \end{aligned}
\end{equation}
In order to prove the invariance of the action \eqref{eqn:actiongenmetric} under generalized diffeomorphisms in the next section, it is very convenient to express it in the form
\begin{equation}\label{eqn:actiongenricci}
  S = \int d^{2D} X\, e^{-2 d}\, \mathcal{R}\,.
\end{equation}
To this end, we rewrite \eqref{eqn:actiongenmetric} as
\begin{equation}
  S = \int d^{2D} X\, e^{-2d}\, \mathcal{R} + \int d^{2D} X \,\sqrt{|H|}\, D_A \Big[ e^{-2\tilde{d}} \big( \nabla_B \mathcal{H}^{AB} - 4 \mathcal{H}^{AB} \nabla_B d \big) \Big]\,,
\end{equation}
where the last term is a vanishing boundary term. Due to the compatibility of the covariant derivative with the generalized vielbein, it is trivial to express the generalized scalar curvature in curved instead of flat indices. One only has to relabel the indices to obtain the desired result \eqref{eqn:gencurvaturecurved} stated in the introduction. The generalized dilaton part of the equation of motion reads
\begin{equation}
  \mathcal{R} =  0\,.
\end{equation}

\subsection{Generalized Ricci tensor}\label{sec:genriccitensor}
Now, we consider the variation of the action \eqref{eqn:actiongenmetric} with respect to the generalized metric $\mathcal{H}^{AB}$. In analogy to \eqref{eqn:variationd}, we consider
\begin{equation}
  \delta S = \int d^{2D} X\, e^{-2d}\, \delta \mathcal{H}^{AB}\, \mathcal{K}_{AB}\,.
\end{equation}
As discussed in \cite{Hohm:2010pp}, the variation $\delta \mathcal{H}^{AB}$ is symmetric and thus it is sufficient to study the symmetric part of $\mathcal{K}_{AB}$ only. Performing the variation explicitly and afterwards symmetrizing $\mathcal{K}_{AB}$ gives rise to
\begin{align}
\label{eqn:Ktensor}
\mathcal{K}_{AB} &= \frac{1}{8} \nabla_A \mathcal{H}_{CD} \nabla_B \mathcal{H}^{CD} - \frac{1}{4} \big[ \nabla_C - 2 (\nabla_C d) \big] \mathcal{H}^{CD} \nabla_D \mathcal{H}_{AB} + 2 \nabla_{(A} \nabla_{B)} d \\ &\,- \nabla_{(A} \mathcal{H}^{CD} \nabla_D \mathcal{H}_{B)C} + \big[ \nabla_D - 2 (\nabla_D d) \big] \big[ \mathcal{H}^{CD} \nabla_{(A} \mathcal{H}_{B)C} + {\mathcal{H}^C}_{(A} \nabla_C {\mathcal{H}^D}_{B)} \big]  \nonumber \\ &\,+ \frac{1}{6} F_{ACD} F_B{}^{CD} \nonumber.
\end{align}
Furthermore, the $O(D,D)$ constraint
\begin{equation}
  \mathcal{H}^{AC} \eta_{CD} \mathcal{H}^{DB} = \eta^{AB}
\end{equation}
has to be preserved under the variation \cite{Hohm:2010pp}. This implies that only a certain projection of $\mathcal{K}_{AB}$ gives rise to the equations of motion. Hence, it is necessary to introduce the projection operators
\begin{equation}
P_{AB} = \frac{1}{2} \big( \eta_{AB} - S_{AB} \big) \quad\text{and}\quad\bar{P}_{AB} = \frac{1}{2} \big( \eta_{AB} + S_{AB} \big)\,,
\end{equation}
which are used to define the generalized Ricci tensor
\begin{equation}\label{eqn:projectiongenricci}
  \mathcal{R}_{AB} = 2 {P_{(A}}^C {\bar{P}_{B)}}^{\;\;D} \mathcal{K}_{CD}\,.
\end{equation}
This projection cancels the term in the last line of \eqref{eqn:Ktensor}. Thus, we find a generalized Ricci tensor whose structure matches the one of toroidal DFT. However, all partial derivatives have to be replaced with covariant ones.

\section{Local symmetries}\label{sec:symmetries}
The CSFT derivation of DFT${}_\mathrm{WZW}$ in \cite{Blumenhagen:2014gva} was very challenging. The recasting of the action and the gauge transformations in section \ref{sec:genmetricformulation} is a good, first indication that everything is consistent: All the different terms with bared and unbared indices integrate nicely into doubled objects. However, a much more important consistency check is the invariance of the action \eqref{eqn:actiongenmetric} under the gauge transformations \eqref{eqn:gendiffHAB&tilded}. If all previous calculations were performed correctly, the CSFT framework guarantees this invariance  up to cubic order in the fields. As we will show in subsection \ref{sec:gendiffinv}, it even holds to all higher orders introduced by the generalized metric formulation. Besides generalized diffeomorphism invariance, the action is also manifestly invariant under 2D-diffeomorphisms, as we prove in subsection \ref{sec:2Ddiff}.

\subsection{Generalized diffeomorphisms}\label{sec:gendiffinv}
It does not matter whether one proves the invariance under gauge transformations for the action \eqref{eqn:actiongenmetric} or \eqref{eqn:actiongenricci}. Both only differ by a vanishing total derivative. We choose the latter one, with the generalized curvature scalar $\mathcal{R}$. Proving its invariance, requires two step: First, we show that $\mathcal{R}$ transforms as a scalar under generalized diffeomorphisms. Second, we consider the remaining term $e^{-2d}$ and show that it transforms as a weight +1 scalar density.

In order to show that the generalized curvature \eqref{eqn:gencurvature} is a scalar under generalized diffeomorphisms, we have to compare its transformation behavior under gauge transformations with the results we expect from generalized diffeomorphisms mediated by the generalized Lie derivative. The failure of a quantity $V$ to transform covariantly under generalized diffeomorphisms reads
\begin{equation}\label{eqn:Deltaxigendiff}
  \Delta_\xi V = \delta_\xi V - \mathcal{L}_\xi V,
\end{equation}
where $\mathcal{L}_\xi$ is the generalized Lie derivative
\begin{equation}\label{eqn:genLieVA}
\mathcal{L}_\xi V^A = \xi^B \nabla_B V^A + \big( \nabla^A \xi_B - \nabla_B \xi^A \big) V^B,
\end{equation}
with the usual generalization to higher rank tensors. $\delta_\xi$ denotes the gauge transformations \eqref{eqn:gendiffHAB&tilded} discussed in section \ref{sec:gaugetrafogenmetric}. From the definition \eqref{eqn:Deltaxigendiff}, it is obvious that
\begin{equation}
  \Delta_\xi \mathcal{H}^{AB} = 0 \quad \text{and} \quad \Delta_\xi \tilde d = 0
\end{equation}
hold. Furthermore, $\Delta_\xi$  is linear and fulfills the product rule
\begin{equation}
\Delta_\xi \big( V W \big) = \big( \Delta_\xi V \big) W + V \big( \Delta_\xi W \big)\,.
\end{equation}
Please note that the gauge transformations $\delta_\xi$ act on the fields $\mathcal{H}^{AB}$ and $\tilde d$ only, whereas the generalized Lie derivative $\mathcal{L}_\xi$ acts on the full tensorial structure. As an instructive example take e.g.
\begin{equation}
\Delta_\xi \big( D_A \mathcal{H}^{BC} \big) = \delta_\xi \big( D_A \mathcal{H}^{BC} \big) - \mathcal{L}_\xi \big( D_A \mathcal{H}^{BC} \big) = D_A \big( \mathcal{L}_\xi \mathcal{H}^{BC} \big) - \mathcal{L}_\xi \big( D_A \mathcal{H}^{BC} \big)\,.
\end{equation} 

We now calculate $\Delta_\xi$ for all sub-terms appearing in the generalized curvature scalar \eqref{eqn:gencurvature}. Finally, we combine these results, using the product rule and the linearity of $\Delta_\xi$ to compute $\Delta_\xi \mathcal{R}$. We begin with
\begin{equation}
  \Delta_\xi \big( \nabla_A d \big) = \Delta_\xi \big( D_A \tilde{d} \big) = - \frac{1}{2} D_A \big( D_D \xi^D \big),
\end{equation}
and since
\begin{equation}
  \mathcal{H}^{MN} \nabla_M \nabla_N d =  H^{MN} D_M D_N \tilde{d}
\end{equation}
holds, we only need to consider
\begin{equation}
\Delta_\xi \big( D_A D_B \tilde{d} \big) =  \big( D_A D_B \xi^D \big) D_D \tilde{d} - \frac{1}{2} D_A D_B \big( D_D \xi^D \big) + {F_{BD}}^{C} \big( D_A \xi^D \big) \big( D_C \tilde{d} \big).
\end{equation}
Furthermore, we obtain
\begin{align}
\Delta_\xi \big( \nabla_A \mathcal{H}^{BC} \big) &= 2 D_A D^{(B} \xi_D \mathcal{H}^{C)D} - 2 D_A D_D \xi^{(B} \mathcal{H}^{C)D} + \frac{2}{3} {F^{(B}}_{AE} H^{C)D} \Big( D^E \xi_D - D_D \xi^E \Big) \nonumber \\ &+\frac{4}{3} {F^{(B}}_{DE} \mathcal{H}^{C)E} D_A \xi^D + \frac{2}{3} {F^{(B}}_{DE} \mathcal{H}^{C)E} D^D \xi_A \nonumber \\ &+\frac{2}{3} {F^D}_{AE} \big( D_D \xi^{(B} \big) \mathcal{H}^{C)E} - \frac{2}{3} {F^D}_{AE} \big( D^{(B} \xi_D \big) H^{C)E}
\end{align}
and
\begin{align}
\Delta_\xi \big( \nabla_A \nabla_B & \mathcal{H}^{AB} \big) = \frac{2}{3} F_{ACE} {F^{E}}_{BD} \xi^C D^B \mathcal{H}^{AD} + \frac{4}{3} F_{ACE} {F^{E}}_{BD} \mathcal{H}^{AB} D^C \xi^D - \frac{1}{3} F_{ACE} \mathcal{H}^{AB} D_B D^C \xi^E \nonumber \\ &\,+\frac{2}{3} F_{ACE} \xi^A D^C D_F \mathcal{H}^{EF} + \frac{10}{3} F_{ACE} \mathcal{H}^{AB} D^C D_B \xi^E + 2 F_{ACE} D^A \xi^C D_D \mathcal{H}^{DE} \nonumber \\ &\,+F_{ACE} D^A \mathcal{H}^{DE} D_D \xi^C - \frac{2}{3} F_{ACE} \xi^A D_D D^C \mathcal{H}^{DE} - D_A D_B \xi^C D_C \mathcal{H}^{AB} \nonumber \\ &\,-2D_A \mathcal{H}^{AB} D_C D_B \xi^C - 2 \mathcal{H}^{AB} D_C D_A D_B \xi^C +\frac{2}{27} F_{ACE} F_{BDF} F^{EDF} \mathcal{H}^{BC} \xi^A\,.
\end{align}
On the right hand side, we canceled all terms of the form
\begin{equation}
  \label{eqn:strongconstflux}
    F_{ABC} \big(D^B \cdot \big) \big(D^C \cdot \big) = \big( D^B \cdot \big) \big( [D_A,D_B] \cdot \big) = 0\,.
\end{equation}
They vanish due to the strong constraint \eqref{eqn:strongconst}. Combining these results, we are finally able to calculate $\Delta_\xi$ of the naive generalized Ricci scalar \eqref{eqn:gencurvature} without the $1/6 F_{ACD} F_B{}^{CD} \mathcal{H}^{AB}$ term. It is denoted as $\tilde{\mathcal R}$ and its failure to transform as a scalar under generalized diffeomorphisms reads
\begin{align}
  \Delta_\xi \mathcal{\tilde{R}} &= \frac{1}{6} \Big(\frac{1}{3} F_{AFH} F_{CGI} {F_{E}}^{HI} \eta_{BD} - \frac{1}{3} F_{ACH} F_{EFI} {F_{G}}^{HI} \eta_{BD}  - F_{ABH} F_{CDF} {F_{EG}}^H \Big) \mathcal{H}^{BC} \mathcal{H}^{DE} \mathcal{H}^{FG} \xi^A \nonumber \\ &\,+\frac{1}{3}F_{ACD} {F^{CD}}_E \mathcal{H}^{AB} D_B \xi^E + \frac{1}{6} F_{ACD} {F^{CD}}_E \mathcal{H}^{AB} D^E \xi_B \nonumber \\ &\,+\frac{1}{6} \Big( F_{IAG} {F^G}_{CD} + F_{CIG} {F^G}_{AD} + F_{ACG} {F^G}_{ID} \Big) \mathcal{H}^{BC} \mathcal{H}^{DE} \xi^A D_E \mathcal{H}^{FI} \nonumber \\ &\,+F_{ACD} D^A \xi_B D^C \mathcal{H}^{BD} - \frac{1}{2} F_{ACD} D^A \mathcal{H}^{BD} D_B \xi^C + F_{ACD} \mathcal{H}^{AB} D^D D^C \xi_B \nonumber \\ &\,-\frac{1}{2} F_{ACD} \mathcal{H}^{AB} \mathcal{H}^{EF} D_F \xi^D D^C \mathcal{H}_{BE} +\frac{1}{2} F_{ACD} \mathcal{H}^{AB} \mathcal{H}^{EF} D^C \xi_E D^D \mathcal{H}_{BF} \label{eqn:deltaxiRlong}\,.
\end{align}
Here, we ordered the terms according to the number of derivatives. All terms with three flat derivatives vanish in the same way as they do for toroidal DFT \cite{Hohm:2010pp}. The third line of \eqref{eqn:deltaxiRlong} vanishes due to the Jacobi identity
\begin{equation}\label{eqn:jacobiid}
  F_{AB}{}^E F_{EC}{}^D + F_{CA}{}^E F_{EB}{}^D + F_{BC}{}^E F_{EA}{}^D = 0\,.
\end{equation}
Additionally, one is able to rewrite the first line as
\begin{equation}
\frac{1}{18} \mathcal{H}^{AB} \xi^G \Big( {F_{EA}}^P \mathcal{H}_{PF} + {F_{FA}}^P \mathcal{H}_{PE} \Big) \Big( {F_B}^{J(E} F_{GHJ} + {F_G}^{J(E} F_{HBJ} + {F_H}^{J(E} F_{BGJ} \Big) \mathcal{H}^{F)H}\,,
\end{equation} 
showing that it is zero due to the Jacobi identity, too. Simplifying the remaining terms in \eqref{eqn:deltaxiRlong}, we make use of the $O(D,D)$ property
\begin{equation}
  \mathcal{H}_{AB} \mathcal{H}^{BC} = \delta_A{}^C \quad \text{and following from it} \quad
  D_D \mathcal{H}_{AB} \mathcal{H}^{BC} = - \mathcal{H}_{AB} D_D \mathcal{H}^{BC}\,,
\end{equation}
which gives rise to
\begin{align}
\Delta_\xi \mathcal{\tilde{R}} &= \frac{1}{3}F_{ACD} {F^{CD}}_E \mathcal{H}^{AB} D_B \xi^E + \frac{1}{6} F_{ACD} {F^{CD}}_E \mathcal{H}^{AB} D^E \xi_B \\ &\,+\frac{1}{2} F_{ACD} D^A \xi_B D^C \mathcal{H}^{BD} + F_{ACD} \mathcal{H}^{AB} D^D D^C \xi_B \nonumber\,.
\end{align}
Further, due to the antisymmetry of the structure coefficients we identify
\begin{align}
F_{ACD} \mathcal{H}^{AB} D^D D^C \xi_B &= \frac{1}{2} F_{ACD} \mathcal{H}^{AB} \big[D^D,D^C\big] \xi_B = \frac{1}{2} F_{ACD} \mathcal{H}^{AB} {F^{DC}}_E D^E \xi_B \nonumber \\ &= -\frac{1}{2} F_{ACD} \mathcal{H}^{AB} {F^{CD}}_E D^E \xi_B
\end{align}
and obtain
\begin{equation}
  \Delta_\xi \mathcal{\tilde{R}} = \frac{1}{3}F_{ACD} {F^{CD}}_E \mathcal{H}^{AB} D_B \xi^E - \frac{1}{3} F_{ACD} {F^{CD}}_E \mathcal{H}^{AB} D^E \xi_B +\frac{1}{2} F_{ACD} D^A \xi_B D^C \mathcal{H}^{BD}\,.
\end{equation}
The last term vanishes under the strong constraint \eqref{eqn:strongconstflux} and
\begin{equation}
  \Delta_\xi \mathcal{\tilde{R}} = \frac{1}{3} \mathcal{H}^{AB} F_{ACD} F_E{}^{CD} \big( D_B \xi^E - D^E \xi_B \big)
  \label{eqn:ricciwithout}
\end{equation}
remains. This non-vanishing failure of $\tilde{\mathcal R}$ to transform like a scalar should be canceled by the term
\begin{align}
  \frac{1}{6} F_{ACD} F_B{}^{CD} \mathcal{H}^{AB}
\end{align}
that we have not taken into account yet. Indeed, $\Delta_\xi$ applied on this term gives rise to
\begin{equation}
  \frac{1}{6} \Delta_\xi \big( F_{ACD} F_B{}^{CD} \mathcal{H}^{AB} \big) =
    -\frac{1}{3} \mathcal{H}^{AB} F_{ACD} F_E{}^{CD} \big( D_B \xi^E - D^E \xi_B \big)
\end{equation}
after remembering $\delta_\xi F_{ABC} = 0$ (gauge transformations act on fluctuations only, but not on background fields \cite{Blumenhagen:2014gva}). Ultimately, we obtain the desired result
\begin{equation}
  \Delta_\xi \mathcal{R} = \Delta_\xi \tilde{\mathcal R} + \frac{1}{6} \Delta_\xi \big( F_{ACD}F_B{}^{CD} \mathcal{H}^{AB}\big) = 0
\end{equation}
which proves that the generalized curvature scalar \ref{eqn:gencurvature} is indeed a scalar under generalized diffeomorphisms.

In addition to $\mathcal{R}$, we have to check the transformation behavior of the factor $e^{-2d}$ in the action \eqref{eqn:actiongenricci}. To this end, we first rewrite the generalized Lie derivative of the dilaton fluctuations \eqref{eqn:gendiffHAB&tilded} in terms of covariant derivatives
\begin{equation}\label{eqn:genLietildedcov}
  \mathcal{L}_\xi \tilde d = \xi^A \nabla_A \tilde d - \frac{1}{2} \nabla_A \xi^A
    = \xi^A D_A \tilde d - \frac{1}{2} D_A \xi^A - \frac{1}{6} F^A{}_{AB} \xi^B\,,
\end{equation}
where the last term vanishes due to the unimodularity of the structure coefficients. Next, we consider
\begin{equation}
  \delta_\xi e^{-2 d} = -2 e^{-2 d} \delta_\xi d = -2 e^{-2 d} \mathcal{L}_\xi \tilde d \,,
\end{equation}
where we take into account that the background field $\bar d$ is not affected by gauge transformations. With $\mathcal{L}_\xi \tilde d$ written in terms of covariant derivatives, it is trivial to switch to curved indices. Doing so and plugging in \eqref{eqn:genLietildedcov}, $\delta_\xi e^{-2 d}$ reads
\begin{align}
  \delta_\xi e^{-2 d} &= \xi^I \partial_I e^{-2 d} + e^{-2 d} ( \nabla_I \xi^I  + \xi^I 2 \partial_I \bar d ) = \xi^I \partial_I e^{-2 d} + e^{-2d} ( \nabla_I \xi^I  - \Gamma_{JI}{}^J \xi^I ) \nonumber \\
  &= \xi^I \partial_I e^{-2d} + e^{-2 d} \partial_I \xi^I 
\end{align}
after identifying
\begin{equation}
  2 \partial_I \bar d = - \Gamma_{JI}{}^J
\end{equation}
as explained in \cite{Blumenhagen:2014gva}. Thus, we see that $e^{-2 d}$ transforms like a scalar density with the weight +1 and the integral over the product $e^{-2 d} \mathcal{R}$, which is equivalent to the action, is invariant.

Besides the action, the generalized Lie derivative \eqref{eqn:genLieVA} transforms covariantly under generalized diffeomorphisms. Indirectly, this property has already been proven by showing the closure of the gauge algebra 
\begin{equation}\label{eqn:closuregaugealg}
  [\mathcal{L}_{\xi_1}, \mathcal{L}_{\xi_2}] V^A = \mathcal{L}_{[\xi_1,\xi_2]_\mathrm{C}} V^A
\end{equation}
in \cite{Blumenhagen:2014gva}. However to make it more explicit, we consider
\begin{equation}
  \Delta_\xi \mathcal{L}_\lambda V^A = \mathcal{L}_\xi ( \mathcal{L}_\lambda V^A ) - \mathcal{L}_{\mathcal{L}_\xi \lambda} V^A - \mathcal{L}_\lambda ( \mathcal{L}_\xi V^A) = 0\,.
\end{equation}
In combination with \eqref{eqn:closuregaugealg} it vanishes
\begin{equation}
  \Delta_\xi \mathcal{L}_\lambda V^A = \mathcal{L}_{[\xi,\lambda]_\mathrm{C}} V^A - 
  \mathcal{L}_{\mathcal{L}_\xi \lambda} V^A = 0
\end{equation}
after rewriting the C-bracket
\begin{equation}
  [\xi, \lambda]^A_\mathrm{C} = \mathcal{L}_\xi \lambda^A - \frac{1}{2} \nabla^A (\xi_B \lambda^B) 
\end{equation}
in terms of the generalized Lie derivative and the trivial gauge transformation $-1/2 \nabla^I (\xi_J \lambda^J)$.

\subsection{2D-diffeomorphisms}\label{sec:2Ddiff}
Besides the generalized diffeomorphisms discussed in the previous subsection, one can change the coordinates parameterizing the fields of DFT${}_\mathrm{WZW}$ through the standard Lie derivative. This gives rise to $2D$-diffeomorphisms under which the action \eqref{eqn:actiongenricci} is even manifestly invariant. In order to prove this claim, we follow very similar steps as in subsection \ref{sec:gendiffinv}. However, in this case we will not apply the strong constraint in any of the following steps.

Again, we start by introducing the failure
\begin{equation}
\Delta_\xi V = \delta_\xi V - L_\xi V\,.
\end{equation}
of an arbitrary quantity $V$ to transform covariantly. Here, we use the standard Lie derivative $L_\xi$ instead of the generalized Lie derivative. The transformation behavior of the generalized vielbein $E_A{}^I$ and the generalized dilaton fluctuations $\tilde d$ is given by
\begin{align}
  \delta_\xi E_A{}^I &= L_\xi E_A{}^I = \xi^J \partial_J E_A{}^I - E_A{}^J \partial_J \xi^I \quad \text{and} \\
  \delta_\xi \tilde{d} &= L_\xi \tilde{d} = \xi^P \delta_P \tilde{d}\,.
\end{align}
From these two equations, we see that  $E_A{}^I$ transforms as a vector and $\tilde d$ as a scalar under 2D-diffeomorphisms. Next, we check the failure
\begin{equation}\label{eqn:DeltaxinablaIVJ}
  \Delta_\xi \big( \nabla_I V^J \big) = \Delta_\xi \big( \partial_I V^J \big) + \Delta_\xi \big( {\Gamma^J}_{IL} \big) V^L
\end{equation}
of the covariant derivative
\begin{equation}
  \nabla_I V^J = \partial_I V^J + \Gamma_{IK}{}^J V^K
\end{equation}
to transform as a covariant quantity. Being called a `covariant' derivative, this failure should vanish of course. We start by calculating the first term in \eqref{eqn:DeltaxinablaIVJ} and obtain
\begin{equation}
  \Delta_\xi \big( \partial_I V^J \big) = - V^K \partial_K \partial_I \xi^J\,.
\end{equation}
The second terms is a bit more challenging. In order to evaluate it, we need the definition of the Christoffel symbols
\begin{equation}
  \Gamma_{IJ}{}^K = -\frac{1}{3} \bigl(2 \Omega_{IJ}{}^K + \Omega_{JI}{}^K \bigr),
\end{equation}
where $\Omega_{IJK}$ denotes the coefficients of anholonomy
\begin{equation}\label{eqn:OmegaIJK}
  \Omega_{IJK} = E^A{}_I E^B{}_J E^C{}_K \Omega_{ABC} = -\partial_I E^A{}_J E_{AK}\,
\end{equation}
in curved indices. With these definitions at hand, one obtains
\begin{equation}\label{eqn:DeltaxiGammaIJK}
  \Delta_\xi \Omega_{IJ}{}^K = - \partial_I \partial_J \xi^K
    \quad \text{and finally} \quad
  \Delta_\xi \Gamma_{IJ}{}^K = \partial_I \partial_J \xi^K\,.
\end{equation}
Thus, \eqref{eqn:DeltaxinablaIVJ} gives rise to the expected result
\begin{equation}
  \Delta_\xi \big( \nabla_I V^J \big) = -V^K \partial_K \partial_I \xi^J + V^K \partial_I \partial_K \xi^J = 0
\end{equation}
and $\nabla_I$ is indeed a covariant derivative under 2D-diffeomorphisms.

Even though we have shown the vanishing $\Delta_\xi$ of a covariant derivative applied on a vector, this result generalizes to arbitrary tensors. Especially, the failures
\begin{equation}
  \Delta_\xi \big( \nabla_I \mathcal{H}^{JK} \big) = 0
    \quad \text{and} \quad
  \Delta_\xi \big( \nabla_I d \big) = \Delta_\xi \big( \partial_I \tilde{d} \big) = 0
\end{equation}
vanish. The last ingredient in the definition of the generalized curvature scalar \eqref{eqn:gencurvature} are the structure coefficients $F_{IJK}$. Fortunately, their failure to transform covariantly
\begin{equation}
  \Delta_\xi F_{IJ}{}^K = 2 \Omega_{[IJ]}{}^K = \partial_{[I} \partial_{J]} \xi^K = 0
\end{equation}
vanishes, too. Applying the linearity and the product rule of $\Delta_\xi$, we immediately obtain
\begin{equation}
  \Delta_\xi \big( e^{-2 \tilde d} \mathcal{R} \big) = 0, 
\end{equation}
which proves that the product $e^{-2 \tilde d} \mathcal{R}$ transforms as a scalar under 2D-diffeomorphisms. For the action \eqref{eqn:actiongenricci} to be invariant, the remaining factor $e^{-2 \bar d}$ has to transform as a weight +1 scalar density. Indeed, we have
\begin{equation}
  e^{-2\bar d} = \sqrt{|H|}
\end{equation}
which exactly transforms in the right way. Hence, the DFT${}_\mathrm{WZW}$ action exhibits a manifest 2D-diffeomorphism invariance.

Containing covariant derivatives only, the generalized Lie derivative \eqref{eqn:genLieVA} transforms covariantly, too. Hence, it fulfills
\begin{equation}
  \Delta_\xi \mathcal{L}_\lambda V^A = 0\,.
\end{equation}
Rewriting this equation, we obtain
\begin{equation}
  \Delta_\xi \mathcal{L}_\lambda V^A = L_\xi (\mathcal{L}_\lambda V^A) - \mathcal{L}_{L_\xi \lambda} V^A -
    \mathcal{L}_\lambda (L_\xi V^A) = 0\,,
\end{equation}
giving rise to the algebra
\begin{equation}
  [L_\xi, \mathcal{L}_\lambda] V^A = \mathcal{L}_{L_\xi \lambda} V^A
\end{equation}
which links 2D-diffeomorphisms and generalized diffeomorphisms. Equipped with this algebra, our theory implements an extension of the DFT gauge algebra proposed by Cederwall \cite{Cederwall:2014kxa,Cederwall:2014opa}. However, there are some important differences we would like to comment on. Cederwall considered a covariant derivative without torsion on an arbitrary pseudo Riemannian manifold in order to define a generalized Lie derivative formally matching the one of DFT${}_\mathrm{WZW}$. Applying the Bianchi identity without torsion
\begin{equation}
  R_{[IJK]}{}^L = 0
\end{equation}
he shows in full generality that the gauge algebra closes. We consider a torsionful covariant derivative on a group manifold, a very special case of a pseudo Riemannian manifold. Interestingly, the Bianchi identity with torsion
\begin{equation}
  R_{[IJK]}{}^L + \nabla_{[I} T^L{}_{JK]} - T^M{}_{[IJ} T^L{}_{K]M} = 
  \frac{2}{9} \big( F_{IJ}{}^M F_{MK}{}^L + F_{KI}{}^M F_{MJ}{}^L + F_{JK}{}^M F_{MI}{}^L \big) = 0
\end{equation}
reproduces on the group manifold the Jacobi identity which we used to show the closure of the DFT${}_\mathrm{WZW}$ gauge algebra and the invariance of the action under generalized diffeomorphisms. Thus, one is inclined to conjecture that the whole formalism presented here is not limited to a group manifold as background but could hold for arbitrary pseudo Riemannian manifolds.

\section{Transition to original DFT}\label{sec:DFTWZWtoDFT}
Assuming a geometric group manifold as background, in this section we study the connection between DFT${}_\mathrm{WZW}$ and the original formulation. A link between them was already conjectured in \cite{Blumenhagen:2014gva}, but no explicit calculation has been provided yet. Now, with the generalized metric formulation available, we prove that under an additional constraint both theories can be identified. For that purpose, first we introduce a distinguished generalized vielbein in subsection \ref{sec:genvielbein}. Afterwards, we discuss an  additional constraint that links the background fields with the fluctuations around it. We call it the extended strong constraint. As subsection \ref{sec:extendedstrongconst} shows, this constraint allows us to identify the covariant fluxes $\mathcal{F}_{ABC}$ of the DFT flux formulation \cite{Geissbuhler:2013uka,Aldazabal:2013sca,Hassler:2014sba} with the structure coefficients $F_{ABC}$ of the group manifold. Applying the extended strong constraint,  in subsection \ref{sec:=gaugeandaction} we prove the equivalence of the gauge transformations and the action in both theories. In this context, we will briefly discuss the background independence of DFT.

\subsection{Appropriate generalized vielbein}\label{sec:genvielbein}
The starting point for the following discussion is a background generalized vielbein $E_A{}^I$ fulfilling the strong constraint of DFT. Due to 2D-diffeomorphism invariance proven in section \ref{sec:2Ddiff}, one is not forced to parameterize it with the left/right moving coordinates $x^i$/$x^{\bar i}$. Instead, we choose the momentum $x^i$ and winding $\tilde x_i$ coordinates which are common in the generalized metric formulation of DFT \cite{Hohm:2010pp}. They give rise to
\begin{equation}
  X^I=(\tilde x_i \,\,\, x^i)\,,\quad
  \partial_I= (\tilde{\partial}^i\,\,\, \partial_i)
    \quad \text{and} \quad
  \eta_{IJ}=\begin{pmatrix}
    0 & \delta^i_j \\
    \delta_i^j & 0
  \end{pmatrix}\,.
\end{equation}
A canonical choice for the vielbein in the DFT flux formulation \cite{Geissbuhler:2013uka,Aldazabal:2013sca,Hassler:2014sba} is
\begin{equation}\label{eqn:EhatAI}
  E_{\hat A}{}^I = \begin{pmatrix}
    e^a{}_i & 0 \\
    - e_a{}^j B_{ji} &
    e_a{}^i
  \end{pmatrix}\,.
\end{equation}
The strong constraint of DFT requires that it only depends on half of the coordinates. Without any loss of generality, we choose $E_{\hat A}{}^I$ to depend on the momentum coordinates $x^i$. Note that  a hat over a doubled index indicates that the $\eta$-metric
\begin{equation}
  \eta_{\hat A\hat B}=\begin{pmatrix}
    0 & \delta_b^a \\
    \delta^b_a & 0
  \end{pmatrix}
  \quad \text{and its inverse is} \quad
  \eta^{\hat A\hat B}=\begin{pmatrix}
    0 & \delta_a^b \\
    \delta^a_b & 0
  \end{pmatrix}
\end{equation}
are used to lower and raise this index. In order to identify this representation of $\eta$ with the diagonal form \eqref{eqn:etaAB} common in DFT${}_\mathrm{WZW}$, we apply the coordinate independent $O(2 D)$ rotation
\begin{equation}\label{eqn:MAhatB}
  M_A{}^{\hat B} = \begin{pmatrix}
    \eta_{ab} & \delta^b_a \\
    -\eta_{\bar a b} & \delta^b_{\bar a}
  \end{pmatrix}
    \quad \text{with} \quad
    M_A{}^{\hat C} M_B{}^{\hat D} \eta_{\hat C\hat D} = \eta_{AB}\,.
\end{equation}
It leaves the background metric invariant and thus yields
\begin{equation}
    M_A{}^{\hat C} M_B{}^{\hat D} S_{\hat C\hat D} = S_{AB}
      \quad \text{with} \quad
    S_{\hat A\hat B} = \begin{pmatrix}
      \eta^{ab} & 0 \\
      0 & \eta_{ab}
    \end{pmatrix}\,.
\end{equation}
Switching to curved indices, $S^{\hat A\hat B}$ gives rise to the generalized metric
\begin{equation}
  H^{IJ} = E_{\hat A}{}^I S^{\hat A\hat B} E_{\hat B}{}^J = 
    \begin{pmatrix} g_{ij} - B_{ik} g^{kl} B_{lj} & B_{ik} g^{kj} \\
      - g^{ik} B_{kj} & g^{ij}
    \end{pmatrix}\,.
\end{equation}
It is important to note that the canonical generalized vielbein \eqref{eqn:EAI} of DFT${}_\mathrm{WZW}$ is not an $O(D,D)$ element, because it gives rise to different representations of the $\eta$-metric in flat and curved indices, namely
\begin{equation}
  E_A{}^I \eta^{AB} E_B{}^J = \eta^{IJ} = 2 \begin{pmatrix} g^{ij} & 0 \\
    0 & - g^{\bar i\bar j}
  \end{pmatrix}\,.
\end{equation}
This is an apparent problem, if one tries to compare DFT${}_\mathrm{WZW}$ and DFT. A short calculation shows that the generalized vielbein defined in \eqref{eqn:EhatAI} fixes this problem. It fulfills the relation
\begin{equation}\label{eqn:EhatAIO(D,D)}
  E_{\hat A}{}^I \eta^{\hat A\hat B} E_{\hat B}{}^J = \eta^{IJ} = \begin{pmatrix} 0 & \delta_i^j \\
      \delta_j^i & 0
    \end{pmatrix}
\end{equation}
and hence is an $O(D,D)$ matrix.

This new generalized vielbein should give rise to the constant structure coefficients
\begin{equation}
  F_{ABC} = 2 \Omega_{[AB]C} \quad \text{with} \quad
  \Omega_{ABC} = E_{A}{}^I \partial_I E_{B}{}^J E_{C J}
\end{equation}
from which the derivation of DFT${}_\mathrm{WZW}$ in \cite{Blumenhagen:2014gva} starts. Unfortunately, this does not work out because the resulting structure coefficients fail to be constant. A way around is to consider the covariant fluxes
\begin{equation}\label{eqn:FandOmega}
  \mathcal{F}_{\hat A\hat B\hat C} = 3 \Omega_{[\hat A\hat B\hat C]}
\end{equation}
instead. Following \cite{Geissbuhler:2013uka} and remembering that the vielbein $e_a{}^i$ and the $B$-field $B_{ij}$ depend on the momentum coordinates $x^i$ only, we obtain
\begin{align}
  {\mathcal F}_{abc} &= - 3 e_a{}^i e_b{}^j e_c{}^k \partial_{[i} B_{jk]} = - H_{abc} = - F_{abc}
  \quad \text{and} \\
  {\mathcal F}^a{}_{bc} &= 2 e_{[b}{}^i \partial_i e_{c]}{}^j e^a{}_j = 2\Omega_{[bc]}{}^a = F^a{}_{bc} \,.
\end{align}
The remaining independent components $\mathcal{F}^{ab}{}_c$ and $\mathcal{F}^{abc}$ vanish. Next, we switch from $\mathcal{F}_{\hat A\hat B\hat C}$ to $\mathcal{F}_{ABC}$ by applying the transformation $M_A{}^{\hat B}$ defined in \eqref{eqn:MAhatB}. Doing so gives rise to
\begin{equation}\label{eqn:FABChat}
  \mathcal{F}_{ABC} = \begin{cases}
    \mathcal{F}_{abc} + \eta_{ad} \mathcal{F}^d{}_{bc} +  \eta_{bd} \mathcal{F}_a{}^d{}_c +  \eta_{cd} \mathcal{F}_{ab}{}^d = 2 F_{abc} \\
    \mathcal{F}_{\bar a b c} - \eta_{\bar a\bar d} \mathcal{F}^{\bar d}{}_{b c} +  \eta_{b d} \mathcal{F}_{\bar a}{}^{d}{}_c +  \eta_{cd} \mathcal{F}_{\bar a b}{}^d = 0 \\
    \mathcal{F}_{\bar a\bar b c} - \eta_{\bar a\bar d} \mathcal{F}^{\bar d}{}_{\bar b c} -  \eta_{\bar b\bar d} \mathcal{F}_{\bar a}{}^{\bar d}{}_c +  \eta_{cd} \mathcal{F}_{\bar a\bar b}{}^d =  -2 F_{\bar a\bar b c} \\
    \mathcal{F}_{\bar a\bar b\bar c} - \eta_{\bar a\bar d} \mathcal{F}^{\bar d}{}_{\bar b\bar c} -  \eta_{\bar b\bar d} \mathcal{F}_{\bar a}{}^{\bar d}{}_{\bar c} -  \eta_{\bar c\bar d} \mathcal{F}_{\bar a\bar b}{}^{\bar d} =  -4 F_{\bar a\bar b\bar c}\,,
  \end{cases}
\end{equation}
which are constant but still do not match the strict left/right separation in the structure coefficients required to formulate DFT${}_\mathrm{WZW}$. However, there is still a way to cure this problem without spoiling the $O(D,D)$ property \eqref{eqn:EhatAIO(D,D)}. To this end, we apply a coordinate dependent $O(D)\times O(D)$ transformation which acts on
\begin{equation}
  E_A{}^I = M_A{}^{\hat B} E_{\hat B}{}^I = \begin{pmatrix}
    e_{a i} + e_a{}^j B_{ji} & e_a{}^i \\
    -e_{a i} + e_a{}^j B_{ji} & e_a{}^i
  \end{pmatrix}
  \quad \text{as} \quad
  {\tilde E}_A{}^I = T_A{}^B(x^i) E_B{}^I\,.
\end{equation}
In the second row of $E_A{}^I$, we drop the bar over the index a of $e_{a i}$ and $e_a{}^i$ respectively to emphasis that, in contrast to \eqref{eqn:EAI}, we use  the left mover vielbein only. It is connected to the one for the right movers by the $O(D)$ transformation
\begin{equation}
  e_{\bar a}{}^i = t_{\bar a}{}^b e_b{}^i
    \quad \text{with} \quad
  t_{\bar a}{}^b = \mathcal{K}(t_{\bar a}, g t^b g^{-1})\,,
\end{equation}
where $\mathcal{K}$ denotes the Killing form
\begin{equation}\label{eqn:killingform}
  \mathcal{K}(x, y) = - \frac{\alpha' k}{2} \frac{\Tr (\adj_x \adj_y )}{2 h^\vee}
    \quad \text{with} \quad
  x,\,y\in \mathfrak{g}\,,
\end{equation}
introduced in \cite{Blumenhagen:2014gva}, and $g$ is the group element parameterized by the coordinates $x^i$. This transformation is embedded into
\begin{equation}\label{eqn:backgroundvielbein}
  T_A{}^B = \begin{pmatrix}
    \delta_a^b & 0\\
    0 & t_{\bar a}{}^b
  \end{pmatrix}
  \quad \text{producing} \quad
  \tilde E_A{}^I = \begin{pmatrix}
    e_{a i} + e_a{}^j B_{ji} & e_a{}^i \\
    -e_{\bar a i} + e_{\bar a}{}^j B_{ji} & e_{\bar a}{}^i
  \end{pmatrix},
\end{equation}
which `recovers' the correct index structure. Due to the coordinate dependence of this transformation, it modifies the coefficients of anholonomy according to
\begin{equation}\label{eqn:Omegatilde}
  {\tilde \Omega}_{ABC} = T_A{}^D T_B{}^E T_C{}^F ( \Omega_{DEF} - E_D{}^I \partial_I T_{HE} T^H{}_F )\,.
\end{equation}
After some algebra and keeping the definition $t_a = - t_{\bar a}$ in mind, we obtain
\begin{equation}
  \partial_i t_{\bar db} t^{\bar d}{}_c = \mathcal{K}([t_b, t_c], t_a) e^a{}_i = e^a{}_i F_{abc}
\end{equation}
and finally
\begin{equation}
  E_A{}^I \partial_I T_{DB} T^D{}_C = 2 E_A{}^I \begin{pmatrix}
    0 & 0 \\ 0 & - \partial_I t_{\bar d b} t^{\bar d}{}_c
  \end{pmatrix} = - 2 \begin{cases}
    F_{a \bar b\bar c} \\
    F_{\bar a\bar b\bar c} \\
    0 & \text{otherwise.}
  \end{cases}
\end{equation}
This result is nice, because it allows us the fix the problems we encountered with the covariant fluxes $\mathcal{F}_{ABC}$ in \eqref{eqn:FABChat}. After proper antisymmetrization of $\tilde \Omega_{ABC}$, the covariant fluxes for the $O(D)\times O(D)$ rotated generalized vielbein ${\tilde E}_A{}^I$ read
\begin{equation}\label{eqn:covfluxesbackground}
  {\tilde{\mathcal F}}_{ABC} = 2 \begin{cases}
    F_{abc} \\
    - F_{\bar a\bar b\bar c} \\
    0 & \text{otherwise}
  \end{cases}
  \quad \text{or in the standard form} \quad
  {\tilde{\mathcal F}}_{AB}{}^C = \begin{cases}
    F_{ab}{}^c \\
    F_{\bar a\bar b}{}^{\bar c} \\
    0 & \text{otherwise.}
  \end{cases}
\end{equation}
They are now compatible with the left/right separation of the structure coefficients in DFT${}_\mathrm{WZW}$. Thus, via \eqref{eqn:backgroundvielbein} we have succeeded to properly embed the 
WZW background into the flux formulation of original DFT.

\subsection{Extended strong constraint}\label{sec:extendedstrongconst}
There is still a small but peculiar difference in the two definitions of the structure coefficients
\begin{equation}
  F_{ABC} = 2 \Omega_{[AB]C}
    \quad \text{and the covariant fluxes} \quad
  \mathcal{F}_{ABC} = 3 \Omega_{[ABC]}\,.
\end{equation}
In order to identify them even so, first note that $\Omega_{ABC}$ is antisymmetric with respect to its last two indices due to $O(D,D)$ property \eqref{eqn:EhatAIO(D,D)}. Thus, we are able to write
\begin{equation}
  \mathcal{F}_{ABC} = \Omega_{ABC} + \Omega_{CAB} + \Omega_{BCA} = F_{ABC} + \Omega_{CAB}\,.
\end{equation}
Moreover, the purpose of $F_{ABC}$ in DFT${}_\mathrm{WZW}$ is to define the commutator relation
\begin{equation}
  [D_A, D_B] = F_{AB}{}^C D_C
\end{equation}
between flat derivatives. Thus, it is sufficient to study
\begin{equation}\label{eqn:mathcalFABCDC}
  \mathcal{F}_{AB}{}^C D_C \,\cdot = F_{AB}{}^C D_C\, \cdot + \left(D^C E_A{}^I\right) E_B{}_I D_C\, \cdot
\end{equation}
where $\cdot$ denotes arbitrary products of fluctuations $\epsilon^{AB}$, $\tilde d$ and the gauge parameter $\xi^A$, which we also consider as a fluctuation. In DFT${}_\mathrm{WZW}$, the strong constraint only acts on these fluctuations, whereas it does not apply for the background or the relation between background and fluctuations. However, we can of course introduce an additional constraint, the so called extended strong constraint
\begin{equation}\label{eqn:extstrongconst}
  D_A b \, D^A f = 0\,,
\end{equation}
linking background fields $b$ with fluctuations $f$. It restricts all valid field configurations in DFT${}_\mathrm{WZW}$ to a particular subset which allows to cancel the last term in \eqref{eqn:mathcalFABCDC} and therefore to identify $\mathcal{F}_{ABC} = F_{ABC}$. Furthermore, it allows to cancel the last term in the strong constraint in curved indices giving rise to
\begin{equation}\label{eqn:samesc}
  ( \partial_I \partial^I - 2\, \partial_I \bar d\, \partial^I ) \cdot = \partial_I \partial^I \cdot = 0\,,
\end{equation}
which is apparently equivalent to the strong constraint in the original DFT 
formulation.

\subsection{Gauge transformations and action}\label{sec:=gaugeandaction}
Using the covariant fluxes $\mathcal{F}_{ABC}$ instead of the structure coefficients $F_{ABC}$, we have to recalculate the Christoffel symbols of the covariant derivative. To this end, we solve the frame compatibility condition
\begin{equation}
  \nabla_A E_B{}^I = D_A E_B{}^I + \frac{1}{3} \mathcal{F}_{BA}{}^C E_C{}^I + E_A{}^K \Gamma_{KJ}{}^I E_B{}^J = 0
\end{equation}
which gives rise to
\begin{equation}\label{eqn:connectioncovfluxes}
  \Gamma_{IJ}{}^K = -\Omega_{IJ}{}^K + \Omega_{[IJL]}\eta^{LK} = \frac{1}{3} ( -2 \Omega_{IJ}{}^K + \Omega^K{}_{IJ} + \Omega_J{}^K{}_I )\,.
\end{equation}
For this connection, the generalized torsion
\begin{equation}
  \mathcal{T}^I{}_{JK} = 2  \Gamma_{[JK]}{}^I + \Gamma^I{}_{[JK]} = 0\,,
\end{equation}
vanishes. The latter  links the C-bracket
\begin{equation}
  [\xi_1, \xi_2]_\mathrm{C}^I = [\xi_1, \xi_2]_\mathrm{DFT,C}^J + \mathcal{T}^I{}_{JK} \xi_1^J \xi_2^K
\end{equation}
of DFT${}_\mathrm{WZW}$ and DFT. Thus, both theories share besides the strong constraint \eqref{eqn:samesc} the same gauge algebra, too. This also holds for the generalized Lie derivative, which can be derived from the C-bracket as
\begin{equation}
  \mathcal{L}_\xi V^I = [\xi, V]^I_\mathrm{C} + \frac{1}{2} \nabla^I (\xi_J V^J) = [\xi, V]^I_\mathrm{DFT,C} + \frac{1}{2} \partial^I (\xi_J V^J) = \mathcal{L}_{\mathrm{DFT,}\xi} V^I \,.
\end{equation}
Even if the Christoffel symbols $\Gamma_{IJ}{}^K$ get modified, they still keep their transformation behavior
\eqref{eqn:DeltaxiGammaIJK} under 2D-diffeomorphisms. In this sense, 2D-diffeomorphisms are still a manifest symmetry of the action and its gauge transformations. However, this symmetry gets partially broken due to the constraint
\begin{equation}
  L_\xi \eta^{IJ} = 0 = \partial^J \xi^I + \partial^I \xi^J
\end{equation}
which preserves the $O(D,D)$ property \eqref{eqn:EhatAIO(D,D)} of the background generalized vielbein $E_A{}^I$. Further, the strong constraint for $E_A{}^I$ and the extended strong constraint have to transform covariantly, which gives rise to the additional restrictions
\begin{align}
  \Delta_\xi ( \partial_I E_A{}^J \partial^I f ) &= -E_A{}^K \partial_K \partial_I \xi^J \partial^I f = 0 \,, \\
  \Delta_\xi ( \partial_I E_A{}^J \partial^I E_B{}^K ) &= - E_A{}^L \partial_L \partial_I \xi^J \partial^I E_B{}^K  - \partial_I E_A{}^J E_B{}^L \partial_L \partial^I \xi^K = 0
\end{align}
requiring
\begin{equation}
  \partial_I \xi^J \partial^I f = 0 \quad \text{and} \quad
  \partial_I \xi^J \partial^I E_A{}^K = 0 \quad \text{or} \quad
  \partial_I \xi^K = \text{const.} \,.
\end{equation}
The latter allows for global $O(D,D)$ rotations. Besides them, only transformations of the form
\begin{equation}
  L_\xi E_A{}^I = \xi^J \partial_J E_A{}^I + E_A^J \partial_J \xi^I = E_A{}^J \begin{pmatrix}
    0 & 0 \\
    \partial_{[j} \tilde \xi_{i]} & 0 \\
  \end{pmatrix} 
\end{equation}
are possible. They correspond to $B$-field gauge transformations with
\begin{equation}
  B_{ij} \to B_{ij} + \partial_{[i} \xi_{j]}
\end{equation}
and, as well as the global $O(D,D)$ rotations, can be expressed in terms of generalized diffeomorphisms. Hence, the additional 2D-diffeomorphism invariance of DFT${}_\mathrm{WZW}$ is completely broken by the extended strong constraint \eqref{eqn:extstrongconst} and the $O(D,D)$ valued background generalized vielbein.

The new connection \eqref{eqn:connectioncovfluxes} has a non-trivial effect on the background dilaton $\bar d$ defined in \eqref{eqn:splitdilaton}, too. To be compatible with integration by parts \cite{Blumenhagen:2014gva}, $\bar d$ has to fulfill
\begin{equation}
  \label{eqn:dilatonomega}
  \Omega^J{}_{JI} + 2 \partial_I \bar d = 0
\end{equation}
after using
\begin{equation}
  \Omega_{IJ}{}^J = \Omega_{JI}{}^J\,, \quad \text{a direct consequence of} \quad
  F_{IJ}{}^J = \Omega_{IJ}{}^J - \Omega_{JI}{}^J = 0\,,
\end{equation}
and the antisymmetry of $\Omega_{IJK}$ in its last two indices.

Subsequently, we show that the action $S$ of DFT${}_\mathrm{WZW}$ in curved indices is equivalent to the traditional DFT action
\begin{align}
  S_\mathrm{DFT} = \int d^{2D} X e^{-2d} \Big( & \frac{1}{8} \mathcal{H}^{KL} \partial_K \mathcal{H}_{IJ} \partial_L \mathcal{H}^{IJ} -\frac{1}{2} \mathcal{H}^{IJ} \partial_{J} \mathcal{H}^{KL} \partial_L \mathcal{H}_{IK} \nonumber \\
  \label{eqn:actionDFT}
  & - 2 \partial_I d \partial_J \mathcal{H}^{IJ} + 4 \mathcal{H}^{IJ} \partial_I d \partial_J d \Big)\,.
\end{align}
Of course,
\begin{equation}
  S = S_\mathrm{DFT}
\end{equation}
only holds under the extended strong constraint \eqref{eqn:extstrongconst}. To prove this identity, we show that 
\begin{equation}
  S - S_\mathrm{DFT} = \int d^{2D} X e^{-2d} \Delta
\end{equation}
vanishes. Expressing all covariant derivatives in terms of partial derivatives and the connection \eqref{eqn:connectioncovfluxes}, $\Delta$ can be simplified to
\begin{align}
  \Delta = & \mathcal{H}^{IJ} \Big( \Omega_{IKL} \Omega^{KL}{}_J - \Omega^K{}_{KI} \Omega^L{}_{LJ} +
    \frac{1}{2} \Omega_{KLI}\Omega^{KL}{}_J \Big) \nonumber \\
    & \quad - \Omega_{IJ}{}^K \partial_K \mathcal{H}^{IJ} + 2 \Omega^K{}_{KI} \mathcal{H}^{IJ} \partial_J \tilde d  - \Omega^K{}_{KI} \partial_J \mathcal{H}^{IJ} + 2 \mathcal{H}^{IJ} \Omega_{IJ}{}^K \partial_K \tilde d\,.
\end{align}
The last term in the first line vanishes under the strong constraint of the background fields. After performing integration by parts analogous to \eqref{eqn:ibp} and splitting the generalized dilaton according to \eqref{eqn:splitdilaton}, one obtains
\begin{align}
  - \Omega_{IJ}{}^K \partial_K \mathcal{H}^{IJ} &= - 2 \mathcal{H}^{IJ} \Omega_{IJ}{}^K \partial_K \tilde d + \mathcal{H}^{IJ} \Omega_{IJ}{}^K \Omega^L{}_{LK} + \partial_K \Omega_{IJ}{}^K \mathcal{H}^{IJ} \quad \text{and} \\
  - \Omega^{K}_{KI} \partial_J \mathcal{H}^{IJ} &= - 2 \Omega^K{}_{KI} \mathcal{H}^{IJ} \partial_J \tilde d + \mathcal{H}^{IJ} \Omega^K{}_{KI} \Omega^L{}_{LJ} + \mathcal{H}^{IJ} \partial_I \Omega^K{}_{KJ}\,.
\end{align}
Here, we also have applied \eqref{eqn:dilatonomega} to get rid of derivatives acting on $\bar d$. After these substitutions, $\Delta$ reads
\begin{equation}
  \Delta = \mathcal{H}^{IJ} \big( \Omega_{IKL} \Omega^{KL}{}_J + \Omega_{IJ}{}^K \Omega^L{}_{LK} + \partial_K \Omega_{IJ}{}^K + \partial_I \Omega^K{}_{KJ} \big)\,.
\end{equation}
Finally, by taking the definition of $\Omega_{IJK}$ \eqref{eqn:OmegaIJK} into account, it is straightforward to show that
\begin{equation}
  \partial_K \Omega_{IJ}{}^K + \partial_I \Omega^K{}_{KJ} = - \Omega_{IJ}{}^K \Omega^L{}_{LK} -  \Omega_{IKL} \Omega^{KL}{}_J
\end{equation}
holds and thus one obtains the desired result
\begin{equation}
  \Delta = 0\,.
\end{equation}

The calculations shown in this subsection generalize in some sense the endeavor of \cite{Hohm:2010jy} to find a background independent version of the cubic DFT action derived in \cite{Hull:2009mi}. The main idea behind those technically challenging calculations in that paper is that `\ldots one can absorb a constant part of the fluctuation field $e_{ij}$ into a change of the background field $E_{ij}$. The dilaton plays no role in the background dependence; \ldots' (\cite{Hohm:2010jy} page six, first paragraph). In our context, we have a similar situation by splitting the generalized metric into
\begin{equation}
  \mathcal{H}^{IJ} = H^{IJ} + h^{IJ}, \quad \text{where} \quad
  h^{IJ} = \epsilon^{IJ} + \frac{1}{2} \epsilon^{IK}H_{KL} \epsilon^{LJ} + \dots,
\end{equation}
i.e. the background field $H^{IJ}$ and the fluctuation field $h^{IJ}$. As opposed to \cite{Hohm:2010jy}, we consider the generalized dilaton \eqref{eqn:splitdilaton}, too. Furthermore, we are not limited to constant background fields, because $H^{IJ}$ is not constant for an arbitrary group manifold. It is only constant for the special case of a torus. For  being a consistent background, it always has to fulfill the field equations of the theory. Still, we were able to reproduce the background independence of ordinary DFT proposed by \cite{Hohm:2010jy}. 

As we have seen, for this background independence we have to impose the extended strong constraint, which rules out any solutions beyond SUGRA. To this extend, DFT${}_\mathrm{WZW}$ possesses the same background independence as DFT but still allows to have a glimpse at physics not covered by SUGRA. Moreover, the derivation in this subsection shows that DFT breaks the 2D-diffeomorphism invariance of DFT${}_\mathrm{WZW}$. Especially in the context of doubled sigma models with manifest 2D-diffeomorphism invariance like e.g. in \cite{Nibbelink:2013zda}, this could be interesting.

\section{Outlook}\label{sec:outlook}
In the course of this paper, we have derived the generalized metric formulation of the DFT${}_\mathrm{WZW}$ action and proven its invariance under generalized diffeomorphisms and 2D-diffeomorphisms. Afterwards, we have shown that our theory contains the original formulation of DFT as a subset. To this end, we have restricted the background vielbein $E_A{}^I$ to be $O(D,D)$ valued and to fulfill the strong constraint of DFT. Furthermore, the so called extended strong constraint has to link background and fluctuations. There is no reason why there should not be consistent solutions outside this subset. They are beyond the scope of SUGRA and could contain new physics. Hence, it is worth to study them. 

In general DFT${}_\mathrm{WZW}$ only needs the closure constraint (CC) for background fields $b$ and the strong constraint (SC) for fluctuations $f$. Depending on how one extends these constraint, the following solutions are accessible:
\begin{center}
  \begin{tabular}{l|cccc|l}
    Theory & CC $b$ & SC $b\,$-$\,b$ & SC $b\,$-$f$ & SC $f$-$f$ & Solutions \\
    \hline\hline
    DFT${}_\mathrm{WZW}$ & \cmark & \xmark & \xmark & \cmark & non-geometric \\
    DFT${}_\mathrm{WZW}$ & \cmark & \cmark & \xmark & \cmark & geometric, non-trivial def. of algebra \\
    DFT & \cmark & \cmark & \cmark & \cmark & geometric, T-dual to SUGRA solution.\\
  \end{tabular}
\end{center}
Besides the most general case giving rise to non-geometric backgrounds, one could drop the extended strong constraint but keeping the strong constraint for the background fields. This choice guarantees that the underlying CFT has a modular invariant partition function but still goes beyond conventional SUGRA. Such solutions could be linked to non-trivial deformations of the Courant algebroid underlying the symmetries of DFT. Some of these deformations are known to give rise to non-commutative deformations of the target space in terms of a Poisson structure \cite{KellWald:2008,Deser:2014wva}. Recently, there has been put much effort into understanding non-commutativity and even non-associativity in gravity theories \cite{Lust:2010iy,Blumenhagen:2010hj,Blumenhagen:2011ph,Bakas:2013jwa,Blumenhagen:2013zpa}. All of them are closely connected to backgrounds with fluxes. Being able to handle such kinds of backgrounds, DFT${}_\mathrm{WZW}$ might be an appropriate tool to push these efforts forward. Another interesting challenge  would be an extension from group manifolds to arbitrary background geometries. To this end, one should follow the observations made at the end of section \ref{sec:gendiffinv}.

\acknowledgments
We would like to thank Ilka Brunner, Olaf Hohm, Stefano Massai, Christoph Mayrhofer, Stefan Groot Nibbelink, Felix Rennecke and Barton Zwiebach for helpful discussions. D.L. likes to thank the theory group of CERN for its hospitality. This work was partially supported by the ERC Advanced Grant ``Strings and Gravity''(Grant.No. 32004) and by the DFG cluster of excellence ``Origin and Structure of the Universe''.

\bibliography{literatur}

\providecommand{\href}[2]{#2}\begingroup\raggedright\begin{thebibliography}{10}

\bibitem{Blumenhagen:2014gva}
R.~Blumenhagen, F.~Haßler, and D.~Lüst, {\it {D}ouble {F}ield {T}heory on
  {G}roup {M}anifolds},  \href{http://xxx.lanl.gov/abs/1410.6374}{{\tt
  arXiv:1410.6374}}.

\bibitem{Siegel:1993th}
W.~Siegel, {\it {S}uperspace duality in low-energy superstrings},  {\em
  Phys.Rev.} {\bf D48} (1993) 2826--2837,
  [\href{http://xxx.lanl.gov/abs/hep-th/9305073}{{\tt hep-th/9305073}}].

\bibitem{Hull:2009mi}
C.~Hull and B.~Zwiebach, {\it {D}ouble {F}ield {T}heory},  {\em JHEP} {\bf
  0909} (2009) 099, [\href{http://xxx.lanl.gov/abs/0904.4664}{{\tt
  arXiv:0904.4664}}].

\bibitem{Hull:2009zb}
C.~Hull and B.~Zwiebach, {\it {T}he {G}auge algebra of double field theory and
  {C}ourant brackets},  {\em JHEP} {\bf 0909} (2009) 090,
  [\href{http://xxx.lanl.gov/abs/0908.1792}{{\tt arXiv:0908.1792}}].

\bibitem{Hohm:2010jy}
O.~Hohm, C.~Hull, and B.~Zwiebach, {\it {B}ackground independent action for
  double field theory},  {\em JHEP} {\bf 1007} (2010) 016,
  [\href{http://xxx.lanl.gov/abs/1003.5027}{{\tt arXiv:1003.5027}}].

\bibitem{Hohm:2011ex}
O.~Hohm and S.~K. Kwak, {\it {D}ouble {F}ield {T}heory {F}ormulation of
  {H}eterotic {S}trings},  {\em JHEP} {\bf 1106} (2011) 096,
  [\href{http://xxx.lanl.gov/abs/1103.2136}{{\tt arXiv:1103.2136}}].

\bibitem{Aldazabal:2013sca}
G.~Aldazabal, D.~Marques, and C.~Nunez, {\it {D}ouble {F}ield {T}heory: {A}
  {P}edagogical {R}eview},  {\em Class.Quant.Grav.} {\bf 30} (2013) 163001,
  [\href{http://xxx.lanl.gov/abs/1305.1907}{{\tt arXiv:1305.1907}}].

\bibitem{Berman:2013eva}
D.~S. Berman and D.~C. Thompson, {\it {D}uality {S}ymmetric {S}tring and
  {M}-{T}heory},  {\em Phys.Rept.} {\bf 566} (2014) 1--60,
  [\href{http://xxx.lanl.gov/abs/1306.2643}{{\tt arXiv:1306.2643}}].

\bibitem{Hohm:2013bwa}
O.~Hohm, D.~Lüst, and B.~Zwiebach, {\it {T}he {S}pacetime of {D}ouble {F}ield
  {T}heory: {R}eview, {R}emarks, and {O}utlook},  {\em Fortsch.Phys.} {\bf 61}
  (2013) 926--966, [\href{http://xxx.lanl.gov/abs/1309.2977}{{\tt
  arXiv:1309.2977}}].

\bibitem{Tseytlin:1990nb}
A.~A. Tseytlin, {\it {D}uality {S}ymmetric {F}ormulation of {S}tring {W}orld
  {S}heet {D}ynamics},  {\em Phys.Lett.} {\bf B242} (1990) 163--174.

\bibitem{Dabholkar:2002sy}
A.~Dabholkar and C.~Hull, {\it {D}uality twists, orbifolds, and fluxes},  {\em
  JHEP} {\bf 0309} (2003) 054,
  [\href{http://xxx.lanl.gov/abs/hep-th/0210209}{{\tt hep-th/0210209}}].

\bibitem{Dabholkar:2005ve}
A.~Dabholkar and C.~Hull, {\it {G}eneralised {T}-duality and non-geometric
  backgrounds},  {\em JHEP} {\bf 0605} (2006) 009,
  [\href{http://xxx.lanl.gov/abs/hep-th/0512005}{{\tt hep-th/0512005}}].

\bibitem{Hull:2006va}
C.~M. Hull, {\it {D}oubled {G}eometry and {T}-{F}olds},  {\em JHEP} {\bf 0707}
  (2007) 080, [\href{http://xxx.lanl.gov/abs/hep-th/0605149}{{\tt
  hep-th/0605149}}].

\bibitem{Hitchin:2004ut}
N.~Hitchin, {\it {G}eneralized {C}alabi-{Y}au manifolds},  {\em
  Quart.J.Math.Oxford Ser.} {\bf 54} (2003) 281--308,
  [\href{http://xxx.lanl.gov/abs/math/0209099}{{\tt math/0209099}}].

\bibitem{Gualtieri:2003dx}
M.~Gualtieri, {\it {G}eneralized complex geometry},
  \href{http://xxx.lanl.gov/abs/math/0401221}{{\tt math/0401221}}.

\bibitem{Hull:2004in}
C.~M. Hull, {\it {A} {G}eometry for non-geometric string backgrounds},  {\em
  JHEP} {\bf 0510} (2005) 065,
  [\href{http://xxx.lanl.gov/abs/hep-th/0406102}{{\tt hep-th/0406102}}].

\bibitem{Hull:2009sg}
C.~Hull and R.~Reid-Edwards, {\it {N}on-geometric backgrounds, doubled geometry
  and generalised {T}-duality},  {\em JHEP} {\bf 0909} (2009) 014,
  [\href{http://xxx.lanl.gov/abs/0902.4032}{{\tt arXiv:0902.4032}}].

\bibitem{Hull:2005hk}
C.~Hull and R.~Reid-Edwards, {\it {F}lux compactifications of string theory on
  twisted tori},  {\em Fortsch.Phys.} {\bf 57} (2009) 862--894,
  [\href{http://xxx.lanl.gov/abs/hep-th/0503114}{{\tt hep-th/0503114}}].

\bibitem{Aldazabal:2011nj}
G.~Aldazabal, W.~Baron, D.~Marques, and C.~Nunez, {\it {T}he effective action
  of {D}ouble {F}ield {T}heory},  {\em JHEP} {\bf 1111} (2011) 052,
  [\href{http://xxx.lanl.gov/abs/1109.0290}{{\tt arXiv:1109.0290}}].

\bibitem{Grana:2012rr}
M.~Grana and D.~Marques, {\it {G}auged {D}ouble {F}ield {T}heory},  {\em JHEP}
  {\bf 1204} (2012) 020, [\href{http://xxx.lanl.gov/abs/1201.2924}{{\tt
  arXiv:1201.2924}}].

\bibitem{Aldazabal:2013mya}
G.~Aldazabal, M.~Graña, D.~Marqués, and J.~Rosabal, {\it {E}xtended geometry
  and gauged maximal supergravity},  {\em JHEP} {\bf 1306} (2013) 046,
  [\href{http://xxx.lanl.gov/abs/1302.5419}{{\tt arXiv:1302.5419}}].

\bibitem{Berman:2013cli}
D.~S. Berman and K.~Lee, {\it {S}upersymmetry for {G}auged {D}ouble {F}ield
  {T}heory and {G}eneralised {S}cherk-{S}chwarz {R}eductions},  {\em
  Nucl.Phys.} {\bf B881} (2014) 369--390,
  [\href{http://xxx.lanl.gov/abs/1305.2747}{{\tt arXiv:1305.2747}}].

\bibitem{Hassler:2014sba}
F.~Haßler and D.~Lüst, {\it {C}onsistent {C}ompactification of {D}ouble {F}ield
  {T}heory on {N}on-geometric {F}lux {B}ackgrounds},  {\em JHEP} {\bf 1405}
  (2014) 085, [\href{http://xxx.lanl.gov/abs/1401.5068}{{\tt
  arXiv:1401.5068}}].

\bibitem{Condeescu:2012sp}
C.~Condeescu, I.~Florakis, and D.~Lüst, {\it {A}symmetric {O}rbifolds,
  {N}on-{G}eometric {F}luxes and {N}on-{C}ommutativity in {C}losed {S}tring
  {T}heory},  {\em JHEP} {\bf 1204} (2012) 121,
  [\href{http://xxx.lanl.gov/abs/1202.6366}{{\tt arXiv:1202.6366}}].

\bibitem{Condeescu:2013yma}
C.~Condeescu, I.~Florakis, C.~Kounnas, and D.~Lüst, {\it {G}auged
  supergravities and non-geometric {Q}/{R}-fluxes from asymmetric orbifold
  {CFT}`s},  {\em JHEP} {\bf 1310} (2013) 057,
  [\href{http://xxx.lanl.gov/abs/1307.0999}{{\tt arXiv:1307.0999}}].

\bibitem{Sfetsos:2009vt}
K.~Sfetsos, K.~Siampos, and D.~C. Thompson, {\it {R}enormalization of {L}orentz
  non-invariant actions and manifest {T}-duality},  {\em Nucl.Phys.} {\bf B827}
  (2010) 545--564, [\href{http://xxx.lanl.gov/abs/0910.1345}{{\tt
  arXiv:0910.1345}}].

\bibitem{Cederwall:2014kxa}
M.~Cederwall, {\it {T}he geometry behind double geometry},
  \href{http://xxx.lanl.gov/abs/1402.2513}{{\tt arXiv:1402.2513}}.

\bibitem{Cederwall:2014opa}
M.~Cederwall, {\it {T}-duality and non-geometric solutions from double
  geometry},  {\em Fortsch.Phys.} {\bf 62} (2014) 942,
  [\href{http://xxx.lanl.gov/abs/1409.4463}{{\tt arXiv:1409.4463}}].

\bibitem{Hohm:2010pp}
O.~Hohm, C.~Hull, and B.~Zwiebach, {\it {G}eneralized metric formulation of
  double field theory},  {\em JHEP} {\bf 1008} (2010) 008,
  [\href{http://xxx.lanl.gov/abs/1006.4823}{{\tt arXiv:1006.4823}}].

\bibitem{Geissbuhler:2011mx}
D.~Geissbuhler, {\it {D}ouble {F}ield {T}heory and {N}=4 {G}auged
  {S}upergravity},  {\em JHEP} {\bf 1111} (2011) 116,
  [\href{http://xxx.lanl.gov/abs/1109.4280}{{\tt arXiv:1109.4280}}].

\bibitem{Geissbuhler:2013uka}
D.~Geissbuhler, D.~Marques, C.~Nunez, and V.~Penas, {\it {E}xploring {D}ouble
  {F}ield {T}heory},  {\em JHEP} {\bf 1306} (2013) 101,
  [\href{http://xxx.lanl.gov/abs/1304.1472}{{\tt arXiv:1304.1472}}].

\bibitem{Schon:2006kz}
J.~Schon and M.~Weidner, {\it {G}auged {N}=4 supergravities},  {\em JHEP} {\bf
  0605} (2006) 034, [\href{http://xxx.lanl.gov/abs/hep-th/0602024}{{\tt
  hep-th/0602024}}].

\bibitem{Nibbelink:2013zda}
S.~Groot~Nibbelink, F.~Kurz, and P.~Patalong, {\it {R}enormalization of a
  {L}orentz invariant doubled worldsheet theory},  {\em JHEP} {\bf 1410} (2014)
  114, [\href{http://xxx.lanl.gov/abs/1308.4418}{{\tt arXiv:1308.4418}}].

\bibitem{KellWald:2008}
F.~Keller and S.~Waldmann, {\it {D}eformation {T}heory of {C}ourant
  {A}lgebroids via the {R}othstein {A}lgebra},
  \href{http://xxx.lanl.gov/abs/0807.0584}{{\tt 0807.0584}}.

\bibitem{Deser:2014wva}
A.~Deser, {\it {S}tar products on graded manifolds and $\alpha'$-corrections to
  {C}ourant algebroids from string theory},
  \href{http://xxx.lanl.gov/abs/1412.5966}{{\tt arXiv:1412.5966}}.

\bibitem{Lust:2010iy}
D.~Lüst, {\it {T}-duality and closed string non-commutative (doubled)
  geometry},  {\em JHEP} {\bf 1012} (2010) 084,
  [\href{http://xxx.lanl.gov/abs/1010.1361}{{\tt arXiv:1010.1361}}].

\bibitem{Blumenhagen:2010hj}
R.~Blumenhagen and E.~Plauschinn, {\it {N}onassociative {G}ravity in {S}tring
  {T}heory?},  {\em J.Phys.} {\bf A44} (2011) 015401,
  [\href{http://xxx.lanl.gov/abs/1010.1263}{{\tt arXiv:1010.1263}}].

\bibitem{Blumenhagen:2011ph}
R.~Blumenhagen, A.~Deser, D.~Lüst, E.~Plauschinn, and F.~Rennecke, {\it
  {N}on-geometric {F}luxes, {A}symmetric {S}trings and {N}onassociative
  {G}eometry},  {\em J.Phys.} {\bf A44} (2011) 385401,
  [\href{http://xxx.lanl.gov/abs/1106.0316}{{\tt arXiv:1106.0316}}].

\bibitem{Bakas:2013jwa}
I.~Bakas and D.~Lüst, {\it 3-{C}ocycles, {N}on-{A}ssociative {S}tar-{P}roducts
  and the {M}agnetic {P}aradigm of ${R}$-{F}lux {S}tring {V}acua},  {\em JHEP}
  {\bf 1401} (2014) 171, [\href{http://xxx.lanl.gov/abs/1309.3172}{{\tt
  arXiv:1309.3172}}].

\bibitem{Blumenhagen:2013zpa}
R.~Blumenhagen, M.~Fuchs, F.~Haßler, D.~Lüst, and R.~Sun, {\it
  {N}on-associative {D}eformations of {G}eometry in {D}ouble {F}ield {T}heory},
   {\em JHEP} {\bf 1404} (2014) 141,
  [\href{http://xxx.lanl.gov/abs/1312.0719}{{\tt arXiv:1312.0719}}].

\end{thebibliography}\endgroup
\bibliographystyle{JHEP}
\end{document}